\documentclass[iop,apj,numberedappendix]{emulateapj}

\usepackage{graphicx}
\usepackage{amssymb,amsmath}
\usepackage{epstopdf}
\newcommand{\Fr}{F_{\rm rad}}
\newcommand{\Fe}{F_{\rm eddy}}
\newcommand{\Fc}{F_{\rm c}}
\newcommand{\delad}{\nabla_{\rm ad}}
\newcommand{\Kzzmax}{K_{zz,{\rm crit}}}

\newcommand{\ps}{_{\rm p}}

\newcommand{\vc}[1]{\mbox{\boldmath{$#1$}}}

\DeclareMathSymbol{\varOmega}{\mathord}{letters}{"0A}
\DeclareMathSymbol{\varSigma}{\mathord}{letters}{"06}
\DeclareMathSymbol{\varPsi}{\mathord}{letters}{"09}

\newcommand{\Eq}[1]{equation\,(\ref{#1})}
\newcommand{\Eqs}[2]{equations (\ref{#1}) and~(\ref{#2})}

\newcommand{\Fig}[1]{Fig.~\ref{#1}}

\shorttitle{Hot Jupiter Atmospheres}
\shortauthors{Youdin \& Mitchell}

\begin{document}

\title{The Mechanical Greenhouse:  Burial of Heat by Turbulence in Hot Jupiter Atmospheres}
\author{Andrew N. Youdin}
\affil{Canadian Institute for Theoretical Astrophysics, University of Toronto,
60 St. George Street, Toronto, Ontario M5S 3H8, Canada}
 
 \author{Jonathan L.\ Mitchell}
 \affil{Dept. of Earth \& Space Sciences and Dept. of Atmospheric and Oceanic Sciences, University of California Los Angeles, 595 Charles Young East Drive, Los Angeles, CA 90095-1567, USA}

\begin{abstract}
The intense irradiation received by hot Jupiters suppresses convection in the outer layers of their atmospheres and lowers their cooling rates.  
``Inflated'' hot Jupiters, i.e., those with anomalously large transit radii, require additional sources of heat or suppressed cooling.  We consider the effect of forced turbulent mixing in the radiative layer, which could be driven by atmospheric circulation or by another mechanism.  Due to stable stratification in the atmosphere, forced turbulence drives a downward flux of heat.   Weak turbulent mixing slows the cooling rate by this process, as if the planet was irradiated more intensely.  Stronger turbulent mixing buries heat into the convective interior, provided the turbulence extends to the radiative-convective boundary.  This inflates the planet until a balance is reached between the heat buried into and radiated from the interior.  
We also include the direct injection of heat due to the dissipation of turbulence or other effects.  Such heating is already known to slow planetary cooling.   We find that dissipation also enhances heat burial from mixing by lowering the threshold for turbulent mixing to drive heat into the interior.  Strong turbulent mixing of heavy molecular species such as TiO may be necessary to explain stratospheric thermal inversions.  We show that the amount of mixing required to loft TiO may overinflate the planet by our mechanism.  This possible refutation of the TiO hypothesis deserves further study.  Our inflation mechanism requires a deep stratified layer that only exists when the absorbed stellar flux greatly exceeds the intrinsic emitted flux.  Thus it would be less effective for more luminous brown dwarfs and for longer period gas giants, including Jupiter and Saturn.

\end{abstract}


\section{Introduction}
Hot Jupiters --- giant planets receiving intense irradiation from their host stars --- are the best characterized class of exoplanets.   Their proximity to their host star yields frequent transits of and occultations by their host star that are observed over a range of wavelengths.  Two remarkable features of hot Jupiters are their inflated radii and the variety of infrared emission signatures, some of which have been interpreted to reveal  stratospheric thermal inversions.

Many hot Jupiters have larger radii than standard cooling models predict, even with the intense irradiation from the host stars included in the radiative transfer.  This implies a mechanism that injects heat into and/or retards the loss of heat from the planets' interiors.   See \citet{FortneyBaraffeMilitzer_09} for a review of proposed mechanisms.   

\citet[hereafter GS02]{GuillotShowman_02} argued that atmospheric winds, driven by intense irradiation, could explain inflated radii.  In their model, the kinetic energy of the winds  dissipates as heat below the penetration depth of starlight.  However the energy need not be deposited into the convective interior (a common misconception).  Dissipating energy in outer radiative layers  suffices to delay planetary contraction.  Turbulence, which the winds can trigger via Kelvin-Helmholz instabilities,   is an efficient mechanism to dissipate kinetic energy  \citep{LiGoodman_10}.   MHD drag is an alternative dissipation mechanism provided weather-layer winds extend to the high-pressure, metallic zone of hydrogen \citep{Perna_etal10,BatyginStevenson_10}.

Thermal inversions, i.e.\ regions where the atmospheric temperature rises with height, may also implicate turbulent mixing in hot Jupiter atmospheres.   Transit spectra of several hot Jupiters have been interpreted as being thermally inverted  \citep{Richardson_etal07, Burrows_etal07, Knutson_etal10}.
These observations appear to confirm the predicion of \citet{Hubeny_etal03} that molecular absorbers, mainly TiO, in the stratosphere could generate inversion layers.  However  vapor phase TiO could rain out of the upper atmosphere if it condenses in cold traps.  Turbulent mixing can counteract this settling.    

 \citet[hereafter S09]{Spiegel_etal09} showed that eddy diffusion coefficients\footnote{Eddy diffusion models small-scale turbulent processes in analogy to molecular diffusion, with tracers fluxed down their mean gradients.} of $K_{zz} \approx 10^7$ --- $10^{11}~{\rm cm^2/s}$ are needed to maintain sufficient stratospheric TiO for thermal inversions.  The range of $K_{zz}$ values accounts for the varying extent of cold traps in planets with different thermal profiles and the size of grains that condense.   S09 argued that the need for strong mixing renders the TiO hypothesis ``problematic," pending improved estimates of $K_{zz}$.  One goal of our study is to determine if large $K_{zz}$ values are energetically problematic.

Sulfur has also been proposed as a high-altitude absorber.  Photochemical models of sulfur abundances \citep{Zahnle_etal09} also include turbulent mixing at a strength $K_{zz} \approx 10^7~{\rm cm^2/s}$, though the dependance on $K_{zz}$ is unclear.  Eddy diffusion is also used in brown dwarf models to explain disequilibrium chemical abundances \citep{GriffithYelle99, HubenyBurrows07}.

Turbulence not only mixes chemical species, it also transports heat.  This paper develops a model that includes this turbulent heat flux in the radiative layers of hot Jupiters.  While convection drives an outward flux of energy, forced turbulence in stably stratified regions drives a downward flux of energy.  This effect is distinct from --- though it accompanies --- the dissipation of turbulence as heat, which we also include.

By altering the flow of energy, we change the cooling and contraction rates of hot Jupiters.   For modest levels of turbulent diffusion, the outward radiative flux  is partially offset by the downward flux of mechanical energy.  This reduces the net cooling flux from the convective interior, which self-consistently pushes the radiative convective boundary (RCB) to higher pressure.\footnote{For simplicity we will describe convectively stable regions as ``radiative," even when we include the transport of heat by both turbulence and radiation.}

\begin{figure}[tb]
\plotone{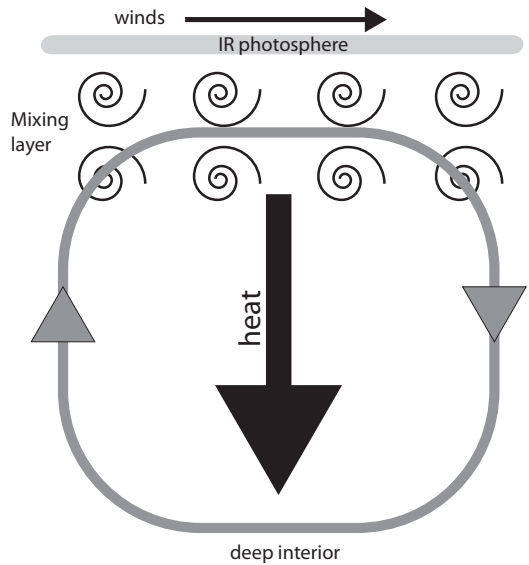}
\caption{Schematic of the mechanical greenhouse effect to inflate hot Jupiters.   A downward flux of heat (\emph{large black arrow}) is driven by turbulence in the convectively stable ``mixing layer" and deposited in the deep interior.  This downward flux can balance or even exceed the convective losses (\emph{gray overturning arrows}).  Atmospheric circulation (\emph{``winds"}) launched near the photosphere drive turbulence in the mixing layer.  Other mechanisms, such as non-linear gravity wave interactions, could also drive the turbulent flux.}
\label{fig:MechGH}
\end{figure}

For sufficiently strong eddy diffusion, the downward flux of energy exceeds the outward radiative flux that a planet of fixed entropy can provide. In this case the turbulent heat flux flows into the convective interior, increasing the internal entropy and inflating the planet.  A schematic of this mechanism is shown in \Fig{fig:MechGH}.  Because higher entropy planets are more intrinsically luminous, inflation leads to an equilibrium between turbulent heat burial and radiative losses.  

Our mechanism bears some resemblance to the runaway greenhouse.  In the latter, the atmosphere is composed of a greenhouse gas in vapor pressure equilibrium with a large, surface volatile reservoir.  The cooling emission to space emanates from a pressure $\sim g/\kappa$, with surface gravity $g$ and (Rosseland mean) opacity $\kappa$.  The emission is independent of the surface temperature for optically thick atmospheres.  Vapor pressure equilibrium determines the temperature at the emission level, thus limiting the cooling radiation that the atmosphere can achieve \citep{Kombayashi_67, Ingersoll_69}.  If the absorbed sunlight exceeds this limiting cooling emission, the surface temperature increases until either the volatile reservoir is depleted or the atmosphere becomes sufficiently transparent to the surface blackbody emission.  The role of limiting cooling flux in the runaway greenhouse is played in our mechanism by the cooling flux of the core.  The role of absorbed sunlight is played by the downward, mechanical flux of energy.  If the latter exceeds the former, the planet heats up by increasing the core entropy until energy balance can be achieved.  The analogy is somewhat incomplete, however, in that the traditional runaway greenhouse involves a radiative-thermodynamic feedback which does not exist in our mechanism.  

This paper is organized as follows.  In \S\ref{sec:model} we review standard radiative equilibrium models.  We add turbulent heat transport and energy injection to our model in \S\ref{sec:nonRE}.  We derive our mixing length formulae for the turbulent transport of heat in \S\ref{sec:Feddy} and present the model equations and our solution methods in \S\ref{sec:tech}.  We provide a prescription relating turbulent diffusion and dissipation in stratified atmospheres in \S\ref{sec:diffdiss}.   We present and analyze our model results in \S\ref{sec:results}.   We first treat constant diffusion (partly to connect with S09) and ignore energy dissipation in \S\ref{sec:constKzz}.  We then add complexity  by considering  a spatially varying $K_{zz}$ in \S\ref{sec:varyKzz} and including energy  dissipation in \S\ref{sec:diffdissres}.   We discuss consequences of changing the opacity law in \S\ref{sec:opacity}.  We compare our results to dynamical simulations of atmospheric circulation in \S\ref{sec:dynamics} and to the TiO diffusion needed for thermal inversions in \S\ref{sec:S09}.  We summarize our results and their implications in \S\ref{sec:concl}.

\section{Standard Atmospheric Models}\label{sec:model}
We start with a review of the standard radiative transfer approximations used in this work  (\S\ref{sec:rt}) and apply them to radiative equilibrium solutions (\S\ref{sec:RE}). We will introduce our notation and parameter choices. \citet[hereafter AB06]{ArrasBildsten06} present a similar analytic model, which they compare to global models with detailed opacities and equation of state (EOS).

\subsection{Radiative Transfer}\label{sec:rt}

Our goal is to understand energy balance.  We focus on the deep atmosphere which is optically thick  both to incoming stellar irradiation and the planet's emitted flux.  Here the equation of radiative diffusion
\begin{equation} 
{dT \over dP}  =  { \Fr \over k_{\rm rad}} \label{eq:TPradeq} 
\end{equation} 
relates the outgoing radiative flux $\Fr$  to the  variation of temperature $T$ with pressure $P$ via the radiative diffusion coefficient
\begin{equation} 
k_{\rm rad} = {16 \sigma T^3 g \over 3  \kappa} \, ,
\end{equation} 
where $\sigma$ is the Stephan-Boltzmann constant.  
Hydrostatic balance, $dP/dz = -\rho g$ allows pressure to replace height $z$ as the vertical corrdinate, with $\rho$ the atmospheric density.  We hold gravity $g$ constant, invoking the plane-parallel approximation for thin atmospheres.

For the Rosseland mean opacity, our calculations will use a power law,
\begin{equation} \label{eq:kappa}
\kappa = \kappa_1 P^\alpha T^{\beta} \equiv \kappa_o \left(P \over P_{\rm kb}\right)^\alpha \left(T \over T_{\rm 2k}\right)^{\beta}    \, .
\end{equation}
The two forms are equivalent, with the constant $\kappa_1$ being more compact, while $\kappa_o$ has units of opacity and is normalized to $P_{\rm kb} = 1$ kbar and $T_{\rm 2k} = 2000$ K.  Unless stated otherwise, calculations will use $\kappa \propto P$, i.e.\ $\alpha = 1, \beta = 0$ as a rough approximation to collision induced molecular opacity.  Our normalization choice of $\kappa_o = 0.18~{\rm cm^2/g}$  will be justified below (\S\ref{sec:RE}).  
We discuss alternate opacity laws in \S\ref{sec:opacity}.
While realistic opacities are only well approximated by power laws over a limited range, this considerable simplification is useful for developing intuition. 

At the top of our atmosphere, we set the temperature to $T_{\rm deep}$, an approach used in AB06 and advocated by \citet{Iro_etal05}.  This approach is valid when the incident stellar flux exceeds the emitted radiation, resulting in a deep isothermal region at the top of the optically thick atmosphere.   In this physical situation, the  precise location of the upper boundary is not important, as we explain further in \S\ref{sec:RE}.   We point the reader to \citet{Hansen08} and \citet{Guillot10} for sophisticated analytic treatments of radiative equilibrium, which are very useful for interpreting computational models.  Note that some \citep{Baraffe_etal03} but not all \citep{SeagarSasselov00} detailed radiative transfer solutions show an extended isotherm in hot Jupiter atmospheres.  One difference between models is the choice of non-gray opacities, which affects the depths at which starlight in different frequency intervals is absorbed. 

The incident stellar flux averaged over the full planetary surface, $F_{\rm irr} \equiv \sigma T_\ast^4$ gives a characteristic temperature
\begin{equation} 
T_\ast \approx 2000~{\rm K} {L_{\ast,\odot}^{1/4} \over M_{\ast,\odot}^{1/6} P_{{\rm day}}^{1/3}}
\end{equation} 
where stellar mass and luminosity, $M_{\ast,\odot}$ and $L_{\ast,\odot}$ are normalized to solar values, and the orbital period $P_{\rm day}$ is normalized to a (24 hour) day.  Horizontal temperature gradients are important for driving winds in the weather layer.  However these winds efficiently smooth temperature gradients at pressures $\gtrsim$ bar, where timescales for advection are shorter than for radiative losses \citep{ShowmanChoMenou09}.  Thus 1D models are appropriate for basic considerations of energy balance. 

Because the planet is not a perfect blackbody,  $T_{\rm deep}$ may not match $T_\ast$.   Greenhouse or anti-greenhouse effects depend on the relative transparency of the atmosphere to stellar and emitted longwave radiation.  (To be clear, we are now referring to standard radiative effects, not the mechanical greenhouse.)  If incoming starlight penetrates below the infrared photosphere, then the greenhouse effect gives $T_{\rm deep} > T_\ast$.  If, however, significant incoming radiation is absorbed above the photosphere, a stratospheric thermal inversion gives $T_{\rm deep} < T_\ast$.    See \citet{Hubeny_etal03} for a more quantitative analysis.  In most of our examples we adopt $T_{\rm deep} = 1500$ K, as appropriate for short ($\sim$ 1 day) orbital periods with a thermal inversion, or for longer periods with no thermal inversion, a greenhouse effect and/or a more luminous host star.

Giant planets, including hot Jupiters, become unstable to convection at depth. 
The lapse rate of the atmosphere
\begin{equation} \label{eq:delrad}
\nabla \equiv {d \ln T \over d \ln P} = {3 \kappa P \over 16 g} {\Fr \over  \sigma T^4}
\end{equation}
characterizes its stability, and the final equality follows from \Eq{eq:TPradeq}.  Convection occurs where $\nabla > \delad$.  We set $\delad = 2/7$, the adiabatic index of  an ideal diatomic gas.  In reality,  non-ideal interactions lower $\delad$ at the high pressures of exoplanet atmospheres, and promote  convection.  At the top the atmosphere, where the optical depth $\tau = \kappa P/g \approx 1$ and $\Fr \ll \sigma T_{\rm deep}^4$, \Eq{eq:delrad} shows that $\nabla \ll 1$ and the atmosphere is indeed stable and nearly isothermal.  For reasonable opacity choices, $\nabla$ increases with depth, and gives a transition to convection (even under the ideal gas approximation).

In convective regions we set $\nabla = \delad$, i.e.\ an adiabatic profile with $T \propto P^{\delad}$.   The efficiency of convective energy transport makes the modest super-adiabaticity negligible.    The level of the adiabat is determined by the internal entropy.  A global calculation of entropy is beyond our illustrative scope.  Instead, motivated by \citet{Hubbard77}, we label our adiabats by $T_1$, the temperature it would have at $P_1 = 1$ bar pressure, even though the adiabat likely does not extend to such low pressure.  We define a reference entropy (per unit mass) $S_{\rm ref}$, corresponding to $T_1 = 250$ K.  Relative entropy values for different $T_1$ are then computed as
\begin{equation} 
\Delta S \equiv S-S_{\rm ref} = C_P \ln(T_1/250~{\rm K})
\end{equation} 
where the specific heat (at constant pressure) $C_P = R/\delad$ is assumed constant. For the gas constant $R = k_B/(\mu m_{\rm p})$ we use a mean molecular weight $\mu = 2.34$ times the proton mass $m_{\rm p}$.

The stable atmosphere matches smoothly onto the convective adiabat at the radiative-convective boundary (hereafter RCB).   Since the temperature $T_{\rm c}$ and pressure $P_{\rm c}$ at the RCB lie on the interior adaibat we require
\begin{equation}  \label{eq:T1}
T_1 = T_{\rm c} \left(P_1\over P_{\rm c}\right)^{\delad}\,  .
\end{equation} 

The location of the RCB is crucial for global evolution.  The secular cooling of the convective interior is determined by the radiative flux, $F_{\rm c}$, leaving the RCB.  Combining equations (\ref{eq:kappa}),  (\ref{eq:delrad}) and then  (\ref{eq:T1}) at the RCB gives
\begin{eqnarray} 
F_{\rm c} &=&{\delad 16 g \sigma T_{\rm c}^{4-\beta} \over 3 \kappa_1 P_{\rm c}^{1+\alpha}} \label{eq:Fc}\\
&=& {\delad 16 g \sigma \over 3 \kappa_1 } \left({T_1 \over P_1^{\delad} P_{\rm c}^{\nabla_\infty - \delad}}\right)^{4-\beta} \propto {T_1^4  \over P_{\rm c}^{6/7}}\, ,
\label{eq:FcAd}
\end{eqnarray} 
with 
\begin{equation} 
\nabla_\infty \equiv (1+\alpha)/(4-\beta) = 1/2\, .
\end{equation} 
Core flux increases with the interior entropy.\footnote{The numerical scaling in \Eq{eq:FcAd} ignores the effect that higher entropy would lower gravity by inflating the planet.  This effect cancels when computing the total luminosity, which is ultimately more important.}   Pushing the RCB to higher pressures decreases the core flux if $\nabla_\infty > \delad$.  This condition is satisfied for our opacity choice, and is generally required for a transition to convection (as we show shortly).   We emphasize that the dependance of $F_{\rm c}$ on $P_{\rm c}$ is independent of the mechanism that changes $P_{\rm c}$, though previous works have mostly considered irradiation.  These basic considerations are useful in interpreting numerical studies of planetary cooling histories and radii evolution \citep{Burrows_etal00, Baraffe_etal03, Chabrier_etal04}.

\subsection{Radiative Equilibrium Solutions}\label{sec:RE}
We now apply the two approximations of radiative equilibrium (RE) to the stable layer.  First, radiation is the only relevant energy transport mechanism.  Thus $\Fr$ in \Eq{eq:TPradeq} is the total flux of energy.  Second the flux is constant through the radiative  layer with $\Fr = F_{\rm c}$, the flux from the convective interior. This assumes that local (thermal) energy release is negligible.  

\begin{figure}[tb]
\plotone{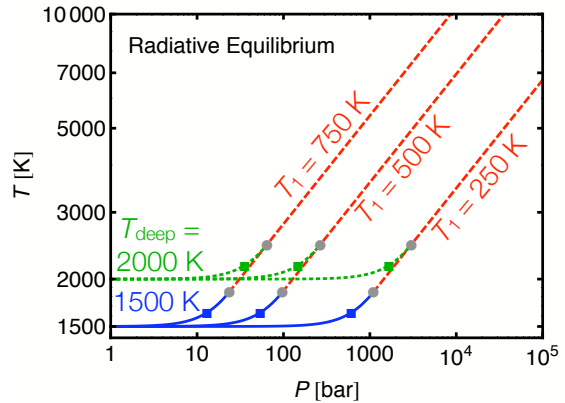}
\caption{Radiative equilibrium (RE) atmospheres with deep isotherms of  $T_{\rm deep} = 1500~{\rm K}$(\emph{blue curves}) and 2000 K (\emph{dotted green curves}) matched onto internal adiabats  (\emph{dashed red curves}) with entropy increasing from bottom to top.  \emph{Grey dots} mark the location ($T_{\rm c}$ and $P_{\rm c}$) of the radiative-convective boundary (RCB), and \emph{squares} show $P_{\rm deep}$, where $\nabla = \delad/2$.}
\label{fig:TPref}
\end{figure}

\Fig{fig:TPref} plots radiative equilbrium atmospheres for $\kappa \propto P$, with two values of $T_{\rm deep}$ matched on to  interior adiabats labeled by $T_1$.   
We obtain analytic RE solutions by integrating  \Eq{eq:TPradeq} with $\Fr = F_{\rm c}$ and $T = T_{\rm deep}$ at $P = 0$ to get
\begin{equation} \label{eq:radeq}
T = T_{\rm deep}\left[1 + {\delad \over \nabla_\infty - \delad} \left({P \over P_{\rm c}}\right)^{1+\alpha}\right] ^{1 /( 4-\beta)}\, .
\end{equation}
This solution uses \Eq{eq:Fc} and we have imposed the requirement $T(P_{\rm c}) = T_{\rm c}$ to find
\begin{eqnarray}
T_{\rm c} &=& T_{\rm deep} \left(\nabla_\infty \over \nabla_\infty - \delad\right)^{1/(4-\beta)} \label{eq:Tc}
\end{eqnarray}
A valid solution --- one that transitions to convection --- thus requires $\nabla_\infty > \delad$ and $\alpha > -1$ (which together  assure $\beta < 4$).   As \Fig{fig:TPref} shows, $T_{\rm c}$  increases with $T_{\rm deep}$ but is independendent of the interior entropy.

The RCB sinks to larger pressure as entropy decreases or as $T_{\rm deep}$ increases, 
\begin{equation}\label{eq:Pc}
P_{\rm c} = k_{\nabla} P_1 \left(T_{\rm deep} \over T_1\right)^{1/\delad}\, ,
\end{equation} 
 which follows from \Eqs{eq:T1}{eq:Tc} with the constant
\begin{equation} 
k_{\nabla} \equiv \left(\nabla_\infty \over \nabla_\infty - \delad \right)^{\nabla_\infty/[\delad(1+\alpha)]} \approx 2.1\,  \nonumber\, .
\end{equation} 

The core flux for RE atmospheres follows from equations (\ref{eq:Fc}), (\ref{eq:Tc}) and (\ref{eq:Pc}) as
\begin{equation} \label{eq:FcRE}
F_{\rm c} = k_F {g \over \kappa_1} \left(T_1 \over  T_{\rm deep}^{1- \delad/\nabla_\infty} \right)^{(1 +\alpha) / \delad} \propto {T_1^{7} \over T_{\rm deep}^3}\, .
\end{equation} 
This gives the well known result that increased irradiation reduces the cooling of the planet, while higher entopy planets are more luminous.  The constant
\begin{equation} 
k_F \equiv {16 \sigma \delad  \over 3 P_1^{1+\alpha}}\left(1 - {\delad \over \nabla_\infty}\right)^{\nabla_\infty/\delad-1}\, . \nonumber
\end{equation} 
Equation (\ref{eq:FcRE}) is consistent with, but more specific than, \Eq{eq:FcAd} in assuming that RE sets the location of $P_{\rm c}$.

We chose the parameters for \Fig{fig:TPref} --- used throughout this work ---  by roughly matching the analytic  solutions to more detailed hot Jupiter models, as in AB06.   

We constrain the entropy parameter $T_1$ by appealing to the typical $P_{\rm c} \approx 1$ kbar location of the RCB in hot Jupiters with modest, i.e.\ Jovian, entropies.  With $T_1 = 260$ K, we reproduce  a 1 kbar RCB for $T_{\rm deep} = 1500$ K.  We also consider larger values of $T_1$ to describe more inflated planets, but keep $P_{\rm c} \gg 1$ bar.

The normalization of the opacity determines the core flux.\footnote{Remarkably, $\kappa_o$ does not affect the location of the RCB along a given adiabat, only $T_{\rm deep}$ and the power laws are required.  Over long times though the opacity at the RCB affects entropy evolution and thereby RCB location.}   Requiring $F_{\rm c} = \sigma (100~{\rm K})^4$ for the standard parameters and $P_{\rm c} = 1$ kbar, gives
$\kappa_{\rm o} = 0.18 {~\rm cm^2/g}$ for $g = 10^3$ cm$^2$/s.  We emphasize that this is not a realistic opacity law (in particular it is too low at small pressures).  We are merely choosing parameters that allow the simple analytic model to mimic properties of more detailed hot Jupiter models.

The lapse rate for RE solutions is (from eq.\ [\ref{eq:radeq}])
\begin{equation} 
\nabla  = \delad { \left(P/P_{\rm c}\right) ^{1+\alpha} \over \left(1-{\delad \over \nabla_\infty}\right) + {\delad \over \nabla_\infty}\left(P/ P_{\rm c}\right)^{1+\alpha}}\, ,
\end{equation}
demonstrating that $\nabla = \delad$ at $P = P_{\rm c}$ and that the solution becomes isothermal $\nabla \rightarrow 0$ at low pressures.  Smooth opacity laws give a monotonic increase in  $\nabla$ with $P$.  Opacity windows give more complicated profiles of $\nabla$, including multiple zones of convection (see \S\ref{sec:opacity}).

We define $P_{\rm deep}$, the effective depth of the isothermal layer, as the location where $\nabla = \delad/2$.  This occurs at
\begin{equation} 
P_{\rm deep} = \left({\nabla_\infty-\delad \over 2 \nabla_\infty-\delad}\right)^{1/(1+\alpha)}P_{\rm c} 
 \approx 0.55 P_{\rm c} \, .
\label{eq:Pdeep}
\end{equation} 
Our definition of $P_{\rm deep}$ differs from AB06, who define  $P_{\rm deep}$ as a characteristic scale that might exceed $P_{\rm c}$.   

We now revisit the validity of applying the boundary condition  $T = T_{\rm deep}$ at $P = 0$.  Due to the isothermal layer at low pressures, applying the boundary condition at any $P \ll P_{\rm deep}$ gives indistinguishable solutions.   However solutions are only physically valid in optically thick regions, for $P \gg P_{\rm thick} \sim g/\kappa_{\rm min} \sim 10~{\rm mbar} [\kappa_{\rm min}/(0.1~{\rm cm^2/g})]^{-1}$.  The relevant opacity $\kappa_{\rm min}$ is the smaller of the opacities to starlight and emitted radiation near $P_{\rm thick}$.  Indeed the penetration of starlight below the infrared photosphere can push the top of the isotherm to $\sim 1$ bar \citep{Guillot10}.   As long as $P_{\rm thick} \ll P_{\rm deep}$, solutions are physically consistent below $P_{\rm thick}$.

\section{Energetics of Turbulent Radiative Layers}\label{sec:nonRE}
We now generalize the radiative equilibrium model to include two effects.  First we allow for turbulent eddies to drive an advective heat flux $\Fe$.  The total flux
\begin{equation} 
F = \Fr + \Fe\, . \label{eq:Ftot} 
\end{equation} 
includes the radiative and eddy contributions.  Second, we allow for the release of energy at a rate $\epsilon$.  In steady state this heating is balanced by cooling from the divergence of the total flux,
\begin{equation} 
{dF \over dP} = - {\epsilon \over g} \label{eq:FP} \, .
\end{equation} 
Sources of $\epsilon$ include the viscous dissipation of turbulence (\S\ref{sec:diffdissres}), breaking waves \citep{ShowmanChoMenou09} and ohmic dissipation \citep{BatyginStevenson_10}.

While \Eq{eq:TPradeq} still describes the radiative flux, $\Fr$ is no longer constant with height.  The fractional contribution of $\Fr$ to the total flux can vary, and the total flux itself can vary.  To proceed further we require a model for $\Fe$ and $\epsilon$.

\subsection{Turbulent Heat Transport}\label{sec:Feddy}
We derive $\Fe$ using basic elements of mixing length theory.  This theory is usually applied to convectively unstable regions, but instead we apply it to forced turbulence in convectively stable regions.  We will show that in this case the energy flux is inwards.  We leave the forcing mechanism of turbulent motions unspecified, and their strength a free parameter.

We consider parcels of gas which conserve entropy and maintain pressure equilibrium with their surroundings as they exchange position over a vertical distance $\ell$ and then dissolve.   These parcels contain  excess heat  $\delta q = \rho C_P \delta T$ with:
\begin{eqnarray}  
\delta T &=& \left(\left. {dT \over dz}\right|_{\rm ad} - {dT \over dz}\right)\ell \nonumber  \\
&=& -{\ell T \over C_P} {dS \over dz}\, 
\end{eqnarray}
where $\delta f$ gives the difference  of any quantity $f$ between the parcel and its surroundings.  For stable stratification ($dS/dz > 0$) rising parcels ($\ell > 0$) cool and sinking parcels heat.  We express the heat flux,  $\Fe = w \delta q$ with $w$ the characteristic eddy speed,  in terms of the turbulent diffusion, $K_{zz} = w \ell$, as
\begin{eqnarray} 
\Fe &=& - K_{zz} \rho T {dS \over dz} \nonumber \\
&=& - K_{zz} \rho g \left(1 - {\nabla \over \delad}\right) \, . \label{eq:Feb}
\end{eqnarray}
The flux is always negative for stable stratification.  It vanishes at the RCB where $d S/dz = 0$ (and $\nabla = \delad$).   We do not model overshoot, which could allow energy exchange (in either direction) between convectively stable and unstable zones.  In the upper isothermal regions $\Fe \propto -K_{zz} P$, declines in magnitude with height, unless $K_{zz}$ increases with height to compensate, as we will consider in \S\ref{sec:varyKzz}.

The above assumption that parcels conserve entropy assumes a radiative cooling time longer than the eddy turnover time.   We are ignoring radiative losses during eddy motion in this work, because we consider optically thick regions with long cooling times.  This assumption will eventually break down in optically thin regions, notably the inversion layer itself, which is strongly stratified.  Whether eddy fluxes affect the structure of the inversion layer is left to future work. 

To understand the energetics of $\Fe$, we analyze its divergence, which describes cooling and (when negative) heating
\begin{equation}\label{eq:Fez}
{d \Fe \over dz} =   K_{zz} {\rho g \over R} {dS \over dz} - K_{zz} \rho T {d^2S \over dz^2} -  {d K_{zz} \over dz}  \rho T {dS \over dz}\, .
\end{equation} 
The terms on the right hand side represent cooling rates, $-\delta \dot{q}$, which we relate to rates of work, $\delta \dot{w}$, using the first law, $\delta q = \delta e - \delta w$.  Thus $-\delta \dot{q} = \delta\dot{w}/\delad$ since the internal energy, $\delta e = \rho C_V \delta T = (1-\delad)\delta q$ with $C_V = C_P - R$.  The first term in  \Eq{eq:Fez} arises from buoyant work $\delta \dot{w}_{\rm B} = \delta \rho g w = K_{zz} \rho g (dS/dz)/C_P$, where $\delta \rho/\rho = -\delta T/T$ by pressure equilibrium.  The first term on the RHS of \Eq{eq:Fez} is thus $-\delta \dot{q}_{\rm B} = \delta\dot{w}_{\rm B}/\delad$.  This buoyant cooling will be evident in the stratified regions ($P < P_{\rm deep}$) of \Fig{fig:Fluxdiff}.

The second term represents the tendency of mixing to heat by filling in entropy minima (for $d^2S /dz^2 > 0$).   More specifically it arises from compressional work, which vanishes for a constant entropy gradient because the work done on rising and sinking parcels cancels.  For varying $dS/dz$ consider two parcels that arrive at $z$, one from above and one from below.   Compressional work is done on the parcels at a rate
\begin{equation} 
 P \nabla \cdot \vc{v}_{\pm} = -{P \over \rho} {\delta \rho_\pm \over \delta t} = \mp{P \ell \over C_P} \left.{dS \over dz}\right|_{ z \pm \ell} {w \over \ell}
\end{equation} 
where the top (bottom) sign refers to sinking (rising) parcels, $\nabla\cdot \vc{v}$ is a velocity divergence, and $\delta t = \ell/w$ gives the expansion rate.  The net work is the sum of these terms, $\delta \dot{w}_{\rm C} = -\delad K_{zz} \rho T d^2S/dz^2$.  We identify the second RHS term in \Eq{eq:Fez} as $-\delta \dot{q}_{\rm C} = \delta\dot{w}_{\rm C}/\delad$.   This compressional heating dominates in deeper regions ($P > P_{\rm deep}$) of \Fig{fig:Fluxdiff}.  The third and final term represents a flux imbalance that arises from non-uniform eddy diffusion as in \S\ref{sec:varyKzz}. 

\subsection{Model Equations and Self-Similar Solution Technique}\label{sec:tech}
Our atmospheric model is described by equations (\ref{eq:TPradeq}), (\ref{eq:Ftot}), (\ref{eq:FP}) and (\ref{eq:Feb}) which reduce to the following  pair of coupled ODEs
\begin{eqnarray} \label{eq:TPed}
{dT \over dP} &=& {F + F_{\rm iso} \over k_{\rm rad}  + {F_{\rm iso} P /( \delad T)}}\\
{d F \over dP} &=& -{\epsilon \over g} \label{eq:FPed}
\end{eqnarray} 
where
\begin{eqnarray} \label{eq:Fiso}
F_{\rm iso} &\equiv&  K_{zz} \rho g
\end{eqnarray} 
is (minus one times) the isothermal limit of the eddy flux.  The thermal profile in \Eq{eq:TPed} describes the combined effects of radiative and eddy fluxes.   Ignoring mixing ($F_{\rm iso} \rightarrow 0$) recovers standard radiative diffusion of \Eq{eq:TPradeq}.  Strong mixing  ($F_{\rm iso} \rightarrow \infty$) creates an isentropic profile ($\nabla \rightarrow \delad$). 

Solving the coupled ODEs for $T$ and $F$ requires prescriptions for $K_{\rm zz}$
and $\epsilon$ and boundary conditions.   As with the RE solutions, we fix $T_{\rm deep}$ at the top of the atmosphere, and match onto an adiabat with a given $T_1$ at the bottom.  This matching generally requires iterative techniques.   
We avoid this complication by finding self-similar solutions.  This technique  is only possible because of the idealizations --- notably power law opacities and ideal gas EOS --- described in \S\ref{sec:rt}.

To find self-similar solutions, we normalize all quantities to their value at the RCB.  As with the analytic RE solutions, we do not know where the RCB is located until we find the solution.  We  integrate outwards from the RCB with trivial boundary conditions: the dimensionless $T$ and $F$ are unity.

We set the strength of diffusion not with a physical value for $K_{zz}$, but by the parameter
 \begin{equation}\label{eq:psi}
\psi_{\rm c}  = {K_{zz}  \rho_{\rm c} g\over   F_{\rm c} }\, ,
\end{equation} 
the ratio of $F_{\rm iso}$ to $F = \Fr$ at the RCB.   For dissipation we must similarly specify  $\epsilon P/(g F)$ at the RCB.  See appendix \ref{sec:appRCBdiss} for details.

To get a physical solution, we scale a dimensionless solution to any desired value of $T_{\rm deep}$ and $T_1$.   Setting $T = T_{\rm deep}$ at $P = 0$, the solution  gives $T_{\rm c}$ at the RCB.  Specification of  $T_1$ then fixes $P_{\rm c}$  via \Eq{eq:T1}.  We then determine $F_{\rm c}$ and $K_{zz}$ via \Eqs{eq:Fc}{eq:psi}.   

\subsection{Relating Diffusion and Dissipation}\label{sec:diffdiss}
Turbulence that gives rise to diffusion, $K_{zz}$, will also dissipate at some rate $\epsilon$.    We now consider a prescription that sets a lower bound on $\epsilon$ from turbulence.   We also allow for stronger heating, perhaps from non-turbulent sources.  

In a Kolmogorov cascade, the dissipation rate $\epsilon = w^3/\ell$ and the diffusion $K_{zz} = w \ell$ give a simple relation $\epsilon = K_{zz}/t_o^2$, with $t_o = \ell/ w$ the turnover time of the dominant eddies.  Unfortunately we lack a reliable model for eddy timescales.  Moreover turbulence in a stratified atmosphere is likely anisotropic and not well described by Kolmogorov scalings.  
Fortunately, our results are not very sensitive to this limitation, as we will show in subsequent sections.
Eddies with long turnover times, $t_o > 1/N$ organize into horizontally extended pancakes, where the squared buoyancy frequency is
\begin{equation} 
N^2 = {g^2\over R T}\left[\delad-\nabla\right]\, .
\end{equation} 

Assuming that the buoyancy frequency sets the relevant timescale, $t_o = 1/N$, gives a dissipation rate
\begin{equation} \label{eq:epbuoy}
\epsilon_{\rm buoy} \approx K_{zz}N^2\, .
\end{equation} 
In strongly stratified, isothermal regions the buoyancy is
\begin{equation} \label{eq:Ndeep}
N_{\rm deep} = g/\sqrt{C_P T_{\rm deep}}  \sim c_{\rm s}/H
\end{equation} 
with $c_{\rm s}$ the sound speed and $H$ the scale height.  For sub-sonic turbulence with $w \ll c_{\rm s}$, our prescription gives $\ell  \sim w/N \ll H$.  This is consistent with the expectation that stratification limits the vertical extent of turbulent structures to less than a scale height.   Forced turbulence with $t_o < 1/N$ is also possible.  Since this would give even stronger dissipation, we consider $\epsilon_{\rm buoy}$ a reasonable lower bound on dissipation for stratified turbulence.

Near the RCB, as $N \rightarrow 0$ it is unreasonable to expect the dissipation to vanish entirely.  Thus we also include a floor to the dissipation
\begin{equation} \label{eq:floor}
\epsilon_o = f_\epsilon K_{zz} N_{\rm deep}^2
\end{equation} 
with the dimensionless normalization $f_\epsilon$ giving the ratio of $\epsilon_o$ to $\epsilon_{\rm buoy}$ in isothermal regions.  If we use rotation as the other relevant timescale, then the floor would be quite low with $f_\epsilon \sim \varOmega^2/N_{\rm deep}^2 \sim 10^{-3} P_{\rm day}^{-7/3}$.  
Our full prescription considers both terms,
\begin{equation}  \label{eq:diss}
\epsilon = \epsilon_o + \epsilon_{\rm buoy}\, ,
\end{equation} 
as discussed in \S\ref{sec:diffdissres}.

\begin{figure}[b]
\plotone{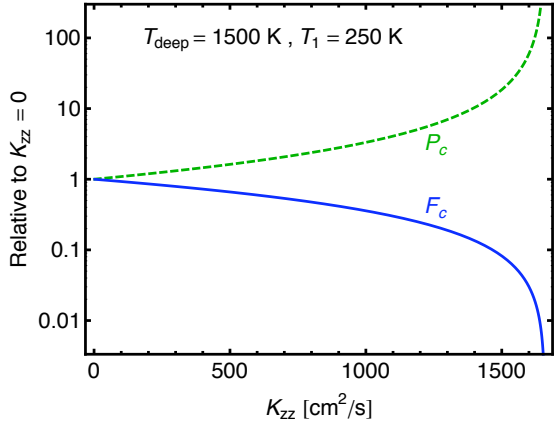}
\caption{As $K_{zz}$ increases the RCB moves to high pressure ($P_{\rm c}$, \emph{dashed green curve}) while the core flux ($F_{\rm c}$, \emph{blue curve}) drops.  Both diverge at at finite $K_{zz} = \Kzzmax \approx 1660~{\rm cm^2/s}$, when dissipation is ignored ($\epsilon =0$).  This upper limit varies with internal entropy and and $T_{\rm deep}$ as shown in \Fig{fig:KzzS}.  Quantities are plotted relative to the $K_{zz}=0$ case where $P_{\rm c} = 1.1$ kbar and $F_{\rm c} = \sigma(100~{\rm K})^4$.
}
\label{fig:Kzzdiff}
\end{figure}

\begin{figure}[tb]
\plotone{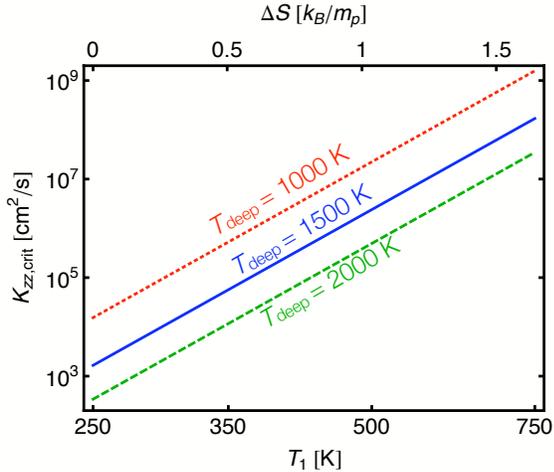}
\caption{Maximum eddy diffusion in the stable layer ($\Kzzmax$), vs.\ internal entropy for $T_{\rm deep} = 1000 , 1500$ and $2000~{\rm K}$ (\emph{dotted red, blue and dashed green curves, respectively}).  See equation (\ref{eq:Kzzmax}) for analytic fits.
Entropy is given in terms of $T_1$ [eq.\ \ref{eq:T1}] and relative to the $T_1 = 250$ K reference (\emph{top axis}).}
\label{fig:KzzS}
\end{figure}

\begin{figure}[t]
\plotone{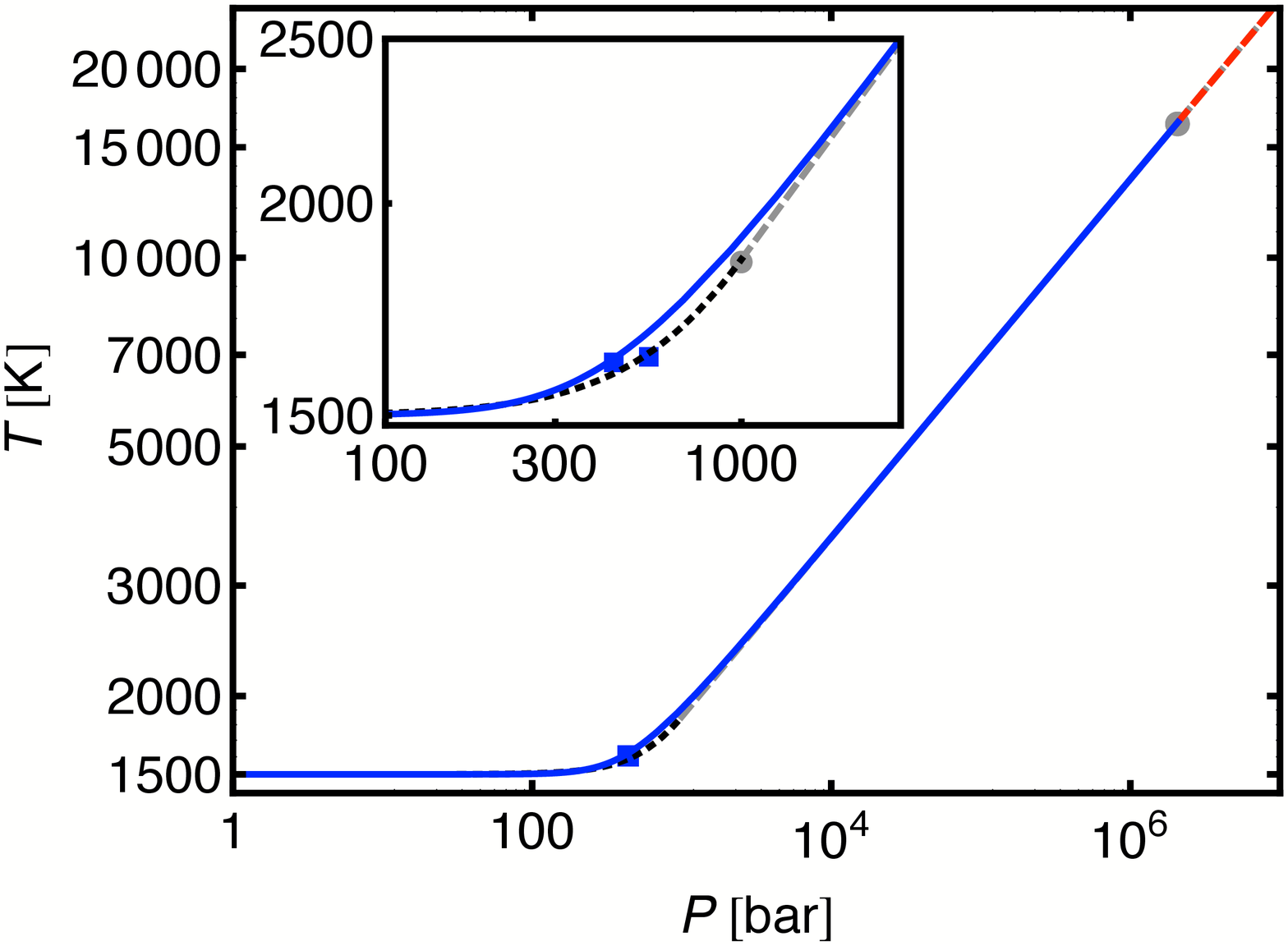}
\plotone{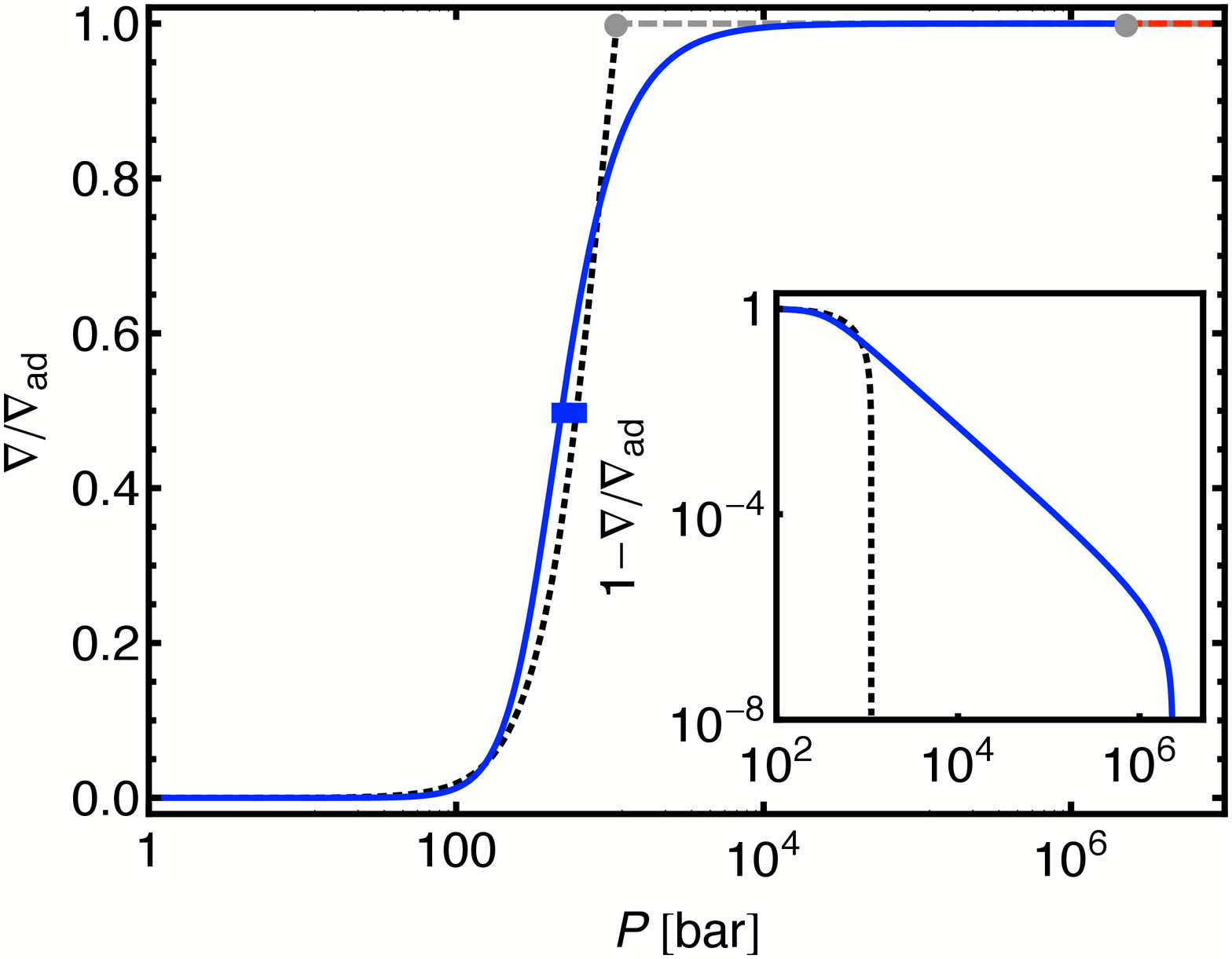}
\caption{Profiles of a stirred atmosphere (\emph{blue curves}) with $K_{zz} \approx \Kzzmax$ and  no dissipation  compared to the RE case with  $K_{zz} = 0$ (\emph{dotted black curves}).  Both join an adiabat with $T_1 = 250$ K (\emph{dashed red and gray curves}).  Mixing pushes the RCB (\emph{gray dots}) to high pressure (which remains finite because $K_{zz}$ is $0.1\%$ below $\Kzzmax$).
(\emph{Left:})  The temperature profile shows modest changes  near $P_{\rm deep}$ (\emph{blue squares}), as shown in the inset.  (\emph{Right:}) Lapse rate $\nabla$, relative to $\delad$.   Turbulent diffusion smoothes the transition towards the adaibat.  The inset plot of $1-\nabla/\delad$ shows the marginal stability of the stirred atmosphere up to high pressures. }
\label{fig:TPdiff}
\end{figure}

\section{Results}\label{sec:results}

\begin{figure}[b]
\plotone{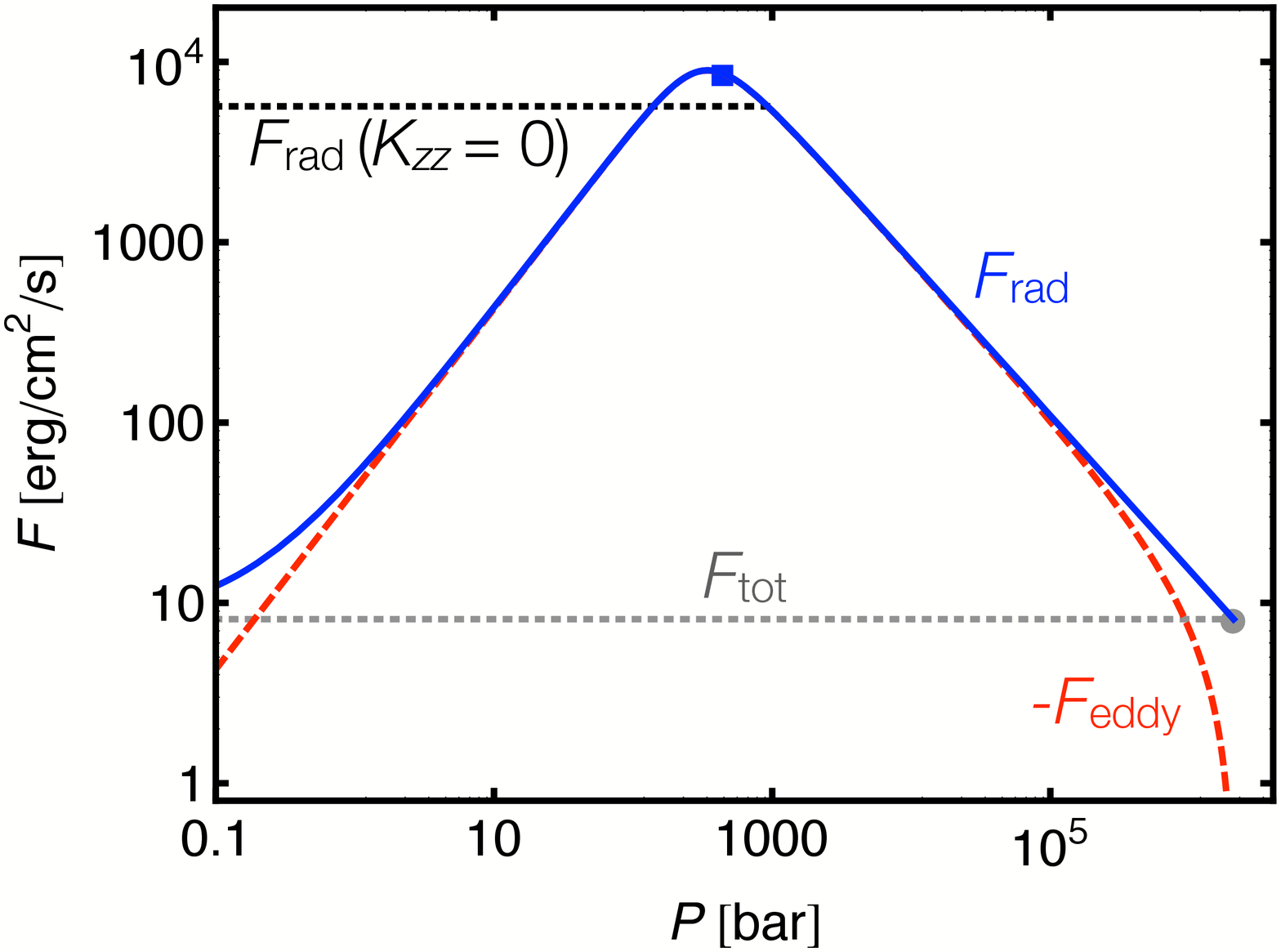}
\caption{Energetic balance for the profiles in \Fig{fig:TPdiff}.  In radiative equilbrium (\emph{black dotted curve}) the radiative flux is constant.   With eddy diffusion and no dissipation the total flux (\emph{gray dotted curve}) is constant.  The radiative flux (\emph{blue curve}) leaving the interior is low, but $\Fr$ first increases and then decreases outward, peaking near $P_{\rm deep}$ (\emph{blue square}).  The downwelling eddy flux (\emph{dashed red curve}) offsets changes to the radiative flux.}
\label{fig:Fluxdiff}
\end{figure}

\subsection{Constant Diffusion, No Dissipation}\label{sec:constKzz}
We describe solutions to the atmopheric model of \S\ref{sec:nonRE}.  We start by considering   turbulent mixing with a constant diffusivity $K_{zz}$ and no dissipation, i.e.\ $\epsilon = 0$.   This requires integration of \Eq{eq:TPed}.   While the decay of turbulence always gives some dissipation, the effect on energetics can be small (as we will show in \S\ref{sec:diffdissres}).  

Our main goal is to constrain the amount of turbulent diffusion that can be maintained in convectively stable regions.  The simplest and most generous --- i.e. allowing the largest levels of turbulent diffusion --- constraint comes when we ignore dissipation.

An appeal to energetic balance shows why this should be the case.   Recalling that turbulence drives a downward eddy flux, $\Fe < 0$, in convectively stable regions, we rearrange flux balance as  $-\Fe = \Fr - F$.  For a large downward eddy flux we need the  total flux $F$ to be small  compared to the outgoing radiative flux. Ignoring dissipation helps in this regard by preventing $F$ from increasing through the layer.  Pushing the RCB to high pressure also helps by lowering the constant $F = F_{\rm c}$ from the RCB.  The solutions below show that strong mixing does indeed push the RCB to high pressure.

\subsubsection{An Upper Limit to $K_{zz}$}
\Fig{fig:Kzzdiff} shows the effect of varying $K_{zz}$ while holding  irradiation ($T_{\rm deep} = 1500$ K) and internal entropy ($T_1 = 250$ K) fixed.  As $K_{zz}$ increases, the downwelling eddy flux pushes the RCB to higher pressures, $P_{\rm c}$, and lowers the flux from the interior, $F_{\rm c}$.  These effects are coupled since $T_{\rm c} \propto P_{\rm c}^{-6/7}$ along a fixed adiabat (eq.\ [\ref{eq:FcAd}]).  Turbulent mixing --- like strong stellar irradiation --- reduces the planet's cooling rate.

At a critical value of  $K_{zz} = \Kzzmax$,  $P_{\rm c}$ diverges to infinity while $F_{\rm c}$ drops to zero.  This upper limit to diffusion is $\Kzzmax \approx 1660~{\rm cm^2/s}$ for the adiabat and $T_{\rm deep}$ chosen in \Fig{fig:Kzzdiff}.   Of course a real planet cannot extend to infinite pressure, to say nothing of the plane-parallel approximation.  The point is that our steady state model cannot energetically support $K_{zz} > \Kzzmax$.  

Stronger turbulence could in principle exist, since we are saying nothing about what forces turbulence.  In a non-equilibrium state with $K_{zz} > \Kzzmax$, the downwelling eddy flux would then increase the internal entropy, and inflate the planet.

\Fig{fig:KzzS} shows that higher entropy planets have a higher  steady state $\Kzzmax$.   Thus by inflating the planet, strong mixing brings the planet's energy balance into equilibrium.  An ultimate upper limit to $K_{\rm zz}$ is that the planet not over-inflate and exceed its observed radius.   The $K_{zz}$ values invoked in S09, from $10^7$ to $> 10^{10} {\rm cm^2/s}$ would imply significant or (on the upper end) excessive inflation.  For comparison AB06 showed (see their Fig. 11) that entropy changes of $\Delta S \approx k_{\rm B}/m_{\rm p}$ (the scale on our \Fig{fig:KzzS}) can expand a hot Jupiter's radius by  $\sim 10$ ---$25\%$.
Accurate determination of the maximum $K_{zz}$ allowed for a given planet requires more detailed modeling (including global structure with realistic opacities and EOS) than we perform.  However our results strongly suggest that $K_{zz}$ values invoked in the literature have significant, or even excessive, effects on energetics.

\Fig{fig:KzzS} also shows that $\Kzzmax$ increases with decreasing $T_{\rm deep}$.  Thus our constraints on mixing are much more stringent for hot Jupiters than for more distant planets, including Jupiter itself.   Recall that thermal inversions lower $T_{\rm deep}$ and that mixing can sustain thermal inversions by keeping opacity sources aloft in the stratosphere.   A planet can accommodate strong mixing with some combination of thermal inversions to lower $T_{\rm deep}$ and increased internal entropy.  It is hard to predict if thermally inverted planets should be more inflated --- due to the presumed presence of turbulence --- or less inflated --- because lower $T_{\rm deep}$ promotes cooling and inhibits our mechanical greenhouse effect.  Observations do not indicate an obvious correlation.  Planets with signatures of inversions exhibit varying degrees of inflation.   See \citet{Miller_etal2009} for a comparison of observed to model radii of transiting planets.

While strong mixing pushes $P_{\rm c} \rightarrow \infty$, the depth of the isothermal layer, $P_{\rm deep}$, is relatively unchanged by mixing.  (We explore structure in detail below.)  Thus planets with higher entropy or lower $T_{\rm deep}$ have shallower isothermal layers.  Specifically $P_{\rm deep} \propto (T_{\rm deep}/T_1)^{1/\delad}$ from \Eqs{eq:Pc}{eq:Pdeep}.  

It is hardly surprising that planets which can accommodate more mixing (larger $\Kzzmax$) have shallower stratified layers to mix (smaller $P_{\rm deep}$).  In principle strong mixing could destroy the deep isotherm altogether by pushing $P_{\rm deep}$ to optically thin regions.  This probably requires unrealistically large core entropies.  Alternatively as interior temperatures rise, blackbody emission at depth may find opacity windows at wavelengths longer than the infrared.  

\subsubsection{Structure and Energetics of Stirred Atmospheres}
We now consider the structure and energetic balance of ``stirred atmospheres" with $K_{zz} \approx \Kzzmax$.  We compare these solutions to standard RE atmospheres of \S\ref{sec:RE} with $K_{zz} = 0$.  Since our model is self-similar, the behavior is independent of the parameters ($T_1$, $T_{\rm deep}$) chosen for illustration.

\Fig{fig:TPdiff} (\emph{top panel}) shows that the  temperature profile of the stirred atmosphere is very similar to the RE case.  In the stirred atmosphere,  the RCB lies at much higher pressure below an extended ``pseudo-adiabat,"\footnote{This is not to be confused with the pseudo-adiabat that describes moist convection in Earth's atmosphere.} which lies very close to the original adiabat.   The inset to \Fig{fig:TPdiff} (\emph{top panel}) focuses on the region near $P_{\rm deep}$ (which drops slightly to $480$ bar from the original 610 bar) where the stirred atmosphere is hotter, by at most $60$ K.   The stirred atmosphere is slightly colder  below 250 bar, though this difference of at most 3 K is not visible.  The stirred atmosphere is very modestly thicker, by $0.06 H_{\rm deep}$, where $H_{\rm deep} = R T_{\rm deep}/g \approx 0.008 R_{\rm J}$.

\Fig{fig:TPdiff} (\emph{bottom panel}) plots the lapse rate. Mixing smoothes the transition towards the adiabat.  The inset shows the smooth decline of $1 - \nabla/\delad$  along the pseudo-adiabat,  which gradually reduces the amplitude of the downwelling eddy flux, from \Eq{eq:Feb}.

\Fig{fig:Fluxdiff} shows that the energetics of the stirred atmosphere differs significantly from the RE case.  With $K_{zz} =0$ there is no eddy flux and the radiative flux is constant down to the RCB, here at $P_{\rm c} \approx 1.1$ kbar.   The stirred atmosphere has a deeper RCB which reduces the core flux significantly.  We could push $P_{\rm c} \rightarrow \infty$ and $F_{\rm c} \rightarrow 0$ with a 0.1\% increase of $K_{zz}$ all the way to $\Kzzmax$, but choose not to for visualization.  

The radiative and eddy fluxes change with height. Their sum --- the total flux --- remains constant  because we ignore dissipation.   \Fig{fig:Fluxdiff} shows that the fluxes behave differently above and below $P_{\rm deep}$, i.e.\ along the isothermal and pseudo-adiabatic regions, respectively.  Along the pseudo-adiabat, the radiative flux declines with depth as $\Fr \propto P^{-6/7}$, as we derived for the core flux in \Eq{eq:Fc}.  The radiative cooling ($d\Fr/dP < 0$) in this region balances heating by eddy diffusion ($d \Fe/dP > 0$).   Achieving this energetic balance requires only modest changes to the $T$-$P$ profile.  Since $\Fe$ scales as $1-\nabla/\delad$, it is very sensitive to small changes in $\nabla$ along the pseudo-adiabat  (see eq.\ [\ref{eq:Feb}] and the bottom inset of \Fig{fig:TPdiff}).

 The energy balance along the isotherm, i.e.\ above $P_{\rm deep}$ is different.  With $\nabla \ll 1$, the eddy flux $\Fe \approx - \rho g K_{zz} \propto -P$ grows in magnitude with depth (a different scaling holds if we vary $K_{zz}$ with height, see \S\ref{sec:varyKzz}).  This localized cooling  ($d \Fe/dP < 0$) balances radiative heating  ($d\Fr/dP >  0$).  The decline in radiative flux with height again requires only modest changes to the thermal profile.  From \Eq{eq:delrad}, $\Fr$ is sensitive to small changes in $\nabla \ll 1$.

We can now simply estimate $\Kzzmax$ using our knowledge that $F_{\rm c} \rightarrow 0$ and that mixing only modestly changes the RE  profile .  
The transition region near $P_{\rm deep}$ is crucial.  Here the eddy flux reaches its peak negative value $F_{\rm eddy,deep} \approx - \rho_{\rm deep}g K_{zz}/2$.   The thermal profile constrains $\Fr$ to be roughly $\Fc(K_{zz}=0)$, the core flux of the RE atmosphere.  We set $F_{\rm rad,deep} \approx 2 \Fc(K_{zz}=0)$ to account for the slightly hotter atmosphere near $P_{\rm deep}$.
Energetic balance, $F_{\rm eddy,deep} + F_{\rm rad,deep} =  \Fc \rightarrow 0$, then gives
\begin{equation}\label{eq:Kzzmax} 
\Kzzmax \approx  {4 \Fc(K_{zz}=0) \over \rho_{\rm deep} g}\propto \left({T_1^{1/ \delad} \over T_{\rm deep}^{1/\delad - 1/\nabla_\infty'}}\right)^{2+\alpha}
\end{equation} 
where $\nabla_\infty' = (2+\alpha)/(5-\beta)$ and our parameters give
$\Kzzmax \propto T_1^{21/2}/T_{\rm deep}^{11/2}$.  These scalings agree with the  results of \Fig{fig:KzzS}.

\subsection{Spatially Varying $K_{zz}$}\label{sec:varyKzz}
The above analysis (\S\ref{sec:constKzz}) shows that upper limits on a constant $K_{zz}$ are set by the balance of eddy and radiative fluxes near $P_{\rm deep}$, which itself scales with internal entropy and $T_{\rm deep}$.  To test the  robustness of this finding,  we include a depth dependence to $K_{zz} \propto P^{\zeta}$.  For winds driven near the photosphere, i.e. the top of our atmospheres, one might expect stronger diffusion in the upper atmosphere, i.e. $\zeta < 0$.   On the other hand if turbulence is triggered by shear layers with the convective interior, perhaps $\zeta > 0$.  As discussed in \S\ref{sec:concl} {\bf [and section 5?]} detailed dynamical simulations can help determine plausible diffusion profiles.

\begin{figure}[tb]
\plotone{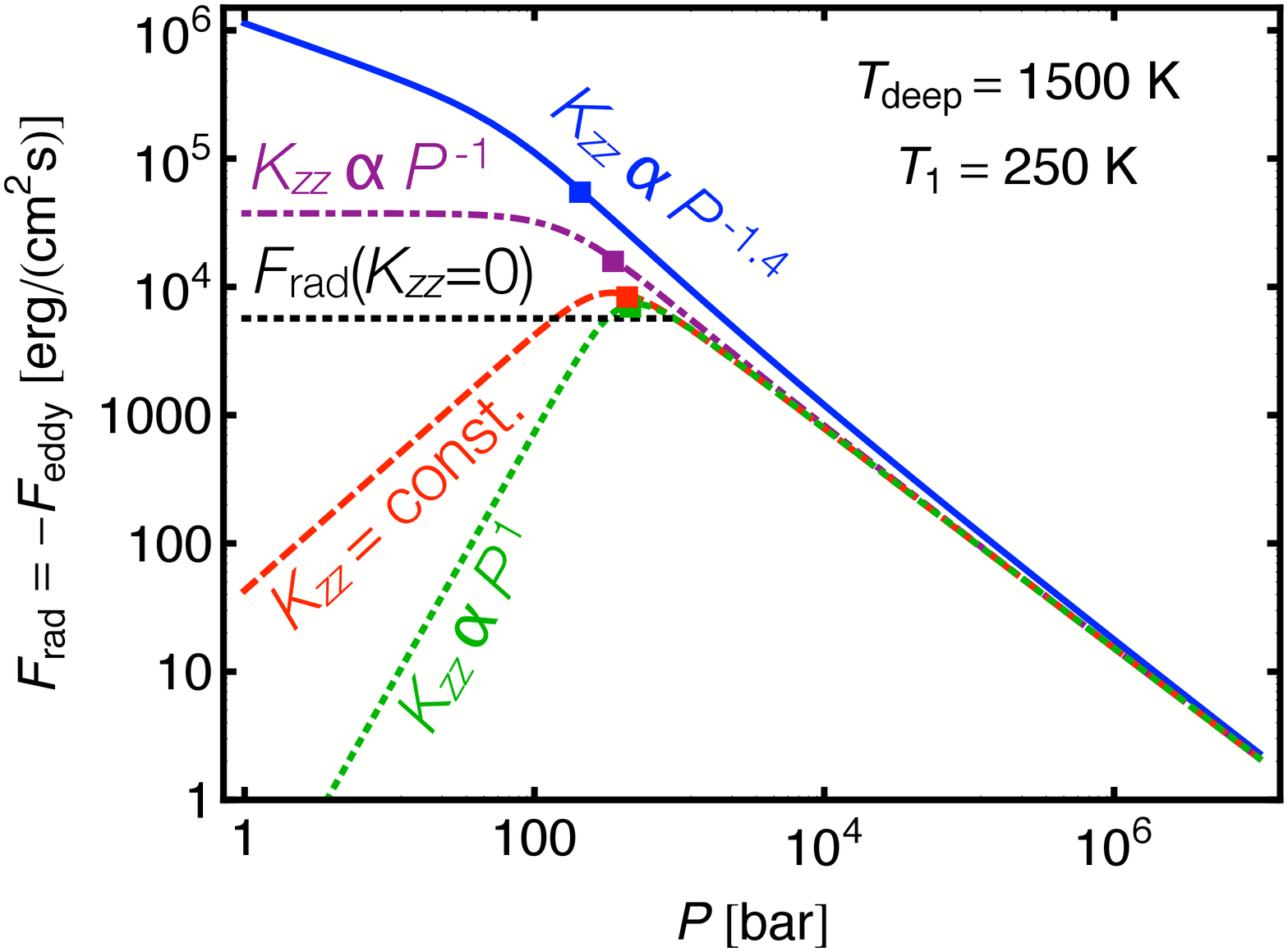}
\plotone{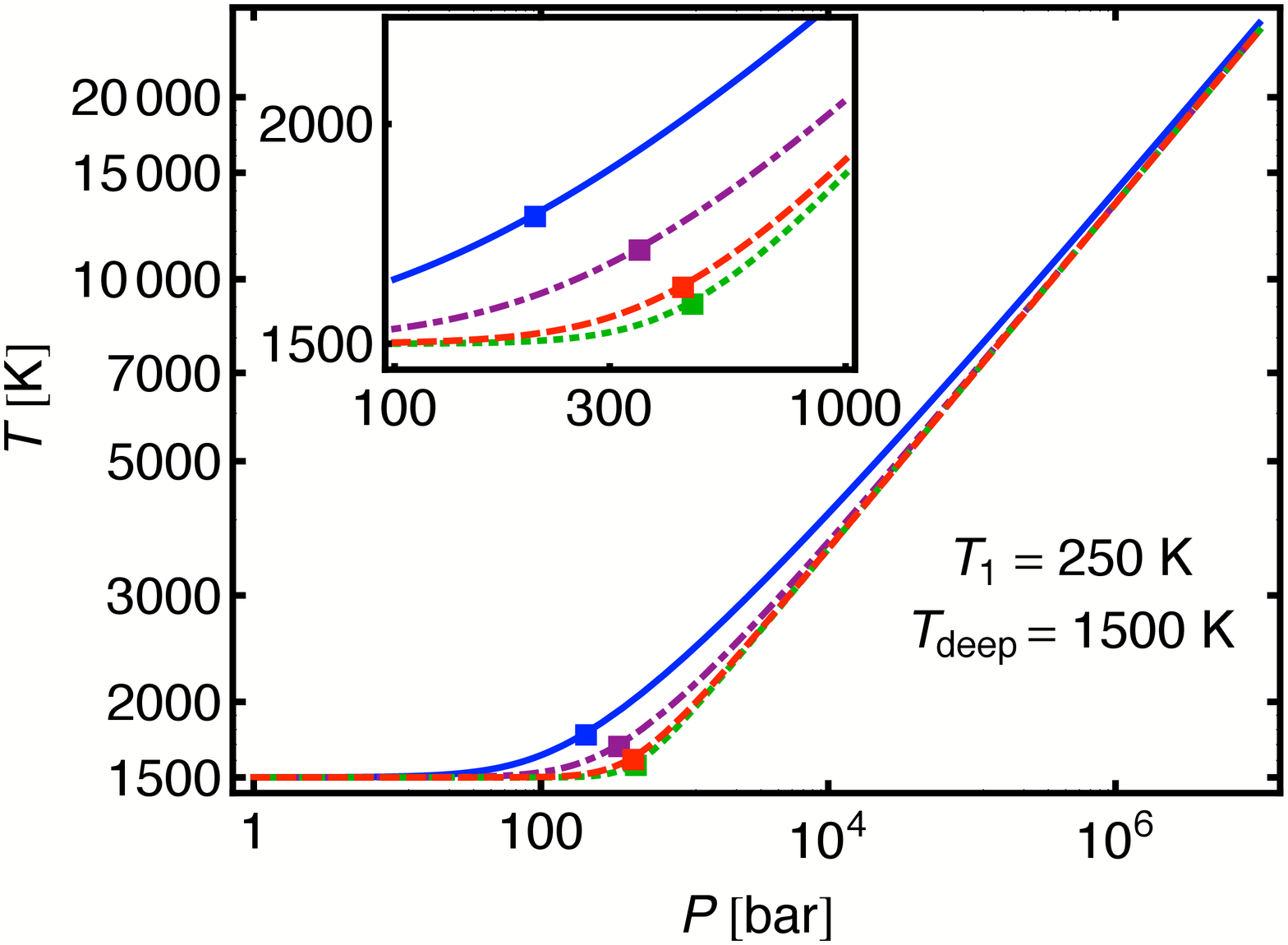}
\caption{Models with a depth-dependent $K_{\rm zz} \propto P^{-\zeta}$ for $\zeta = 1,0,-1$ and $-1.4$(\emph{dotted green, dashed red, dot-dashed purple and solid blue curves, repspectively}) and no dissipation.   Strong mixing pushes $P_{\rm c} \rightarrow \infty$ while $P_{\rm deep}$ is plotted with colored squares.
(\emph{Top}:) Radiative flux, which is equal and opposite to the eddy flux.  Lower $\zeta$ values give strong mixing and larger fluxes at the top of the atmosphere.
(\emph{Bottom}:) Thermal profiles show that strong upper atmosphere mixing (low $\zeta$) heats the upper atmosphere and results in a more gradual approach to the adiabat.}
\label{fig:Kzzvary}
\end{figure}

\Fig{fig:Kzzvary}  shows the effect of varying $\zeta$ with other parameters fixed  (at our standard values of $T_{\rm deep} = 1500$ K, $T_1 = 250$ K, $\alpha = 1$, $\beta=0$ as in e.g.\ \Fig{fig:Kzzdiff}).   These plots show the strongest possible mixing, which (as we found for constant $K_{zz}$) drives the RCB to infinite depths and reduces the core flux to zero.

The maximum mixing near $P_{\rm deep}$ is relatively unchanged, except when the mixing at the top of the atmosphere is quite strong.  Quantitatively we compare  values of $K_{zz,{\rm deep}}$, defined as the maximum value of $K_{zz}$ at  a reference $P = 550$ bar, which is $P_{\rm deep}$ of the radiative equilibrium atmosphere.   For constant $K_{zz}$ we found $\Kzzmax =K_{zz,{\rm deep}} = 1665~{\rm cm^2/s}$.  For mixing that increases with depth as $\zeta = 0.5$, $1.0$, and $1.5$, $K_{zz,{\rm deep}}$ declines by a modest $5\%$, $6\%$ and $5\%$, respectively.
When mixing declines with depth as $\zeta = -0.5$, $-1.0$, and $-1.4$, $K_{zz,{\rm deep}}$ increases by  $14\%$, $58\%$ and $300\%$, respectively.  
We cannot consider models with $\zeta \la -1.5$ because they do not approach an adiabat at depth.
The bottom panel of \Fig{fig:Kzzvary} shows that the approach to the adiabat is already quite gradual for $\zeta = -1.4$.  We explore this issue further in appendix \ref{sec:zeroflux}.  

The top panel of \Fig{fig:Kzzvary} shows the flux profiles for several $\zeta$ values.  The plot shows both radiative and eddy fluxes, which obey $\Fr = -\Fe$ because the net flux, $F \rightarrow 0$ when mixing pushes the RCB to infinite depth.  We also plot the radiative flux for the reference radiative equilibrium model (\emph{horizontal black dotted line}) without any mixing.  The explanation for these flux profiles mirrors the discussion of \Fig{fig:Fluxdiff} in \S\ref{sec:constKzz}.  At high pressures the flux is controlled by the radiative flux along the adiabat.  Increasing or decreasing the mixing with depth has little effect on the deep eddy flux.   Changes to $K_{zz}$ are compensated by $(1 - \nabla/\delad)$ --- see \Eq{eq:Feb} --- which is small and sensitive to slight changes in $\nabla$ close to the adiabat.  

The flux in the low pressure, isothermal region scales as $-\Fe \propto \rho K_{zz} \propto P^{1+\zeta}$ (see eq.\ [\ref{eq:Feb}]).  This explains why the flux components, $\Fr = -\Fe$, increase with height if $\zeta < -1$.   Driving larger radiative fluxes  in the upper atmosphere requires a steeper $dT/dP$.  The temperature profiles in \Fig{fig:Kzzvary} (\emph{bottom panel})  reflect this.  The $\zeta = -1$ and especially the top $\zeta = -1.4$ curves are noticeably hotter at intermediate pressures and have smaller $P_{\rm deep}$.  We thus find that mixing at the top of the atmosphere is more effective --- compared to uniform or bottom-focused mixing --- at lifting (i.e.\ heating) the $T-P$ profile.  The additional heat in this case is provided by a downward flux of mechanical energy across the top boundary.  

\subsubsection{Limits on Mixing Near the Photosphere}\label{sec:Kzztop}
 We now consider what might constrain $K_{zz}$ near the top of the atmosphere, since we find that internal entropy mostly constrains diffusion near $P_{\rm deep}$.   Our $\zeta \leq -1$ solutions in \Fig{fig:Kzzvary} show that strong mixing at the top requires a large  flux of mechanical energy at the top of the atmosphere.   Weather-layer winds are a plausible source of mechanical energy, and they are generated by the atmospheric heat engine driven by insolation.  The thermodynamic efficiency of all planetary atmospheres in the Solar System is of order 1\%.  If we restrict the magnitude of $\Fe$ to a fraction $f_\ast \sim 1\%$ of the insolation $F_\ast \sim \sigma T_{\rm deep}^4$ we get
\begin{eqnarray}\label{eq:Carnot} 
K_{zz,{\rm top}} &<& {f_\ast F_\ast \over \rho_{\rm top} g} \\
&\approx&  10^9~{\rm cm^2 \over s}\left(P_{\rm top} \over 0.1~{\rm bar}\right)^{-1} \left(T_{\rm deep} \over 1500~{\rm K}\right)^5{f_\ast \over 1\%} \, ,\nonumber
\end{eqnarray} 
scaled for a downwelling flux that originates at a 0.1 bar photosphere.  While the efficiency $f_\ast$ and mechanisms of generating a downward mechanical flux are uncertain, the energetic difficulties of mixing at $K_{zz} \gg 10^9~{\rm cm^2/s}$ is evident.

Alternatively, we could attempt to constraint  mixing in the upper atmosphere by appealing to our $\zeta = -1.4$ solution.  This gives the largest downward eddy flux (\Fig{fig:Kzzvary}, \emph{top}).  Smaller $\zeta$ values would be needed for a larger flux, but these do no give consistent solutions.  In  appendix \ref{sec:zeroflux} we analyze the $\zeta \approx -1.4$ limit and argue that it may not be physically significant.   Ignoring this concern, the diffusion near the top of our $\zeta = -1.4$ solutions is $K_{zz} \approx 5 \cdot 10^7 (P/{\rm bar})^{-1.4}~{\rm cm^2/s}$.   A larger internal entropy could support a larger $K_{zz}$ (as we showed for  constant $K_{zz}$ models in \Fig{fig:KzzS}).  Therefore this constraint is not inconsistent with \Eq{eq:Carnot}, which is a more robust constraint.

\begin{figure}[tbh]
\plotone{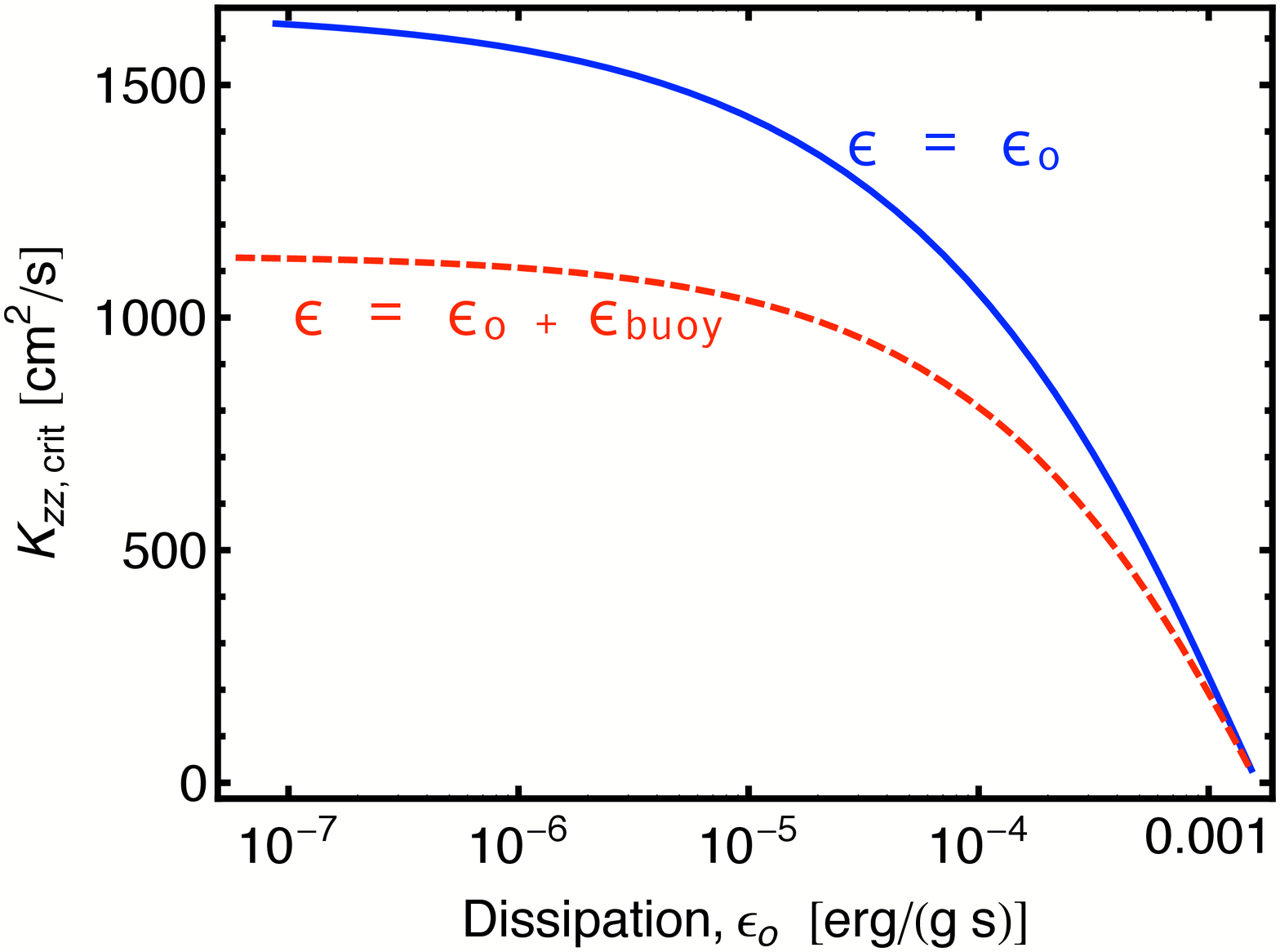}
\caption{Increasing dissipation reduces $\Kzzmax$, shown for fixed $T_{\rm deep} = 1500$ K and $T_1 = 250$ K.   The solid blue curve only includes a constant floor to the dissipation, $\epsilon_o$, while the dashed curve also includes our prescription for dissipation in stratified regions.}
\label{fig:Kzzdiss}
\end{figure}

\begin{figure}[tbh]
\plotone{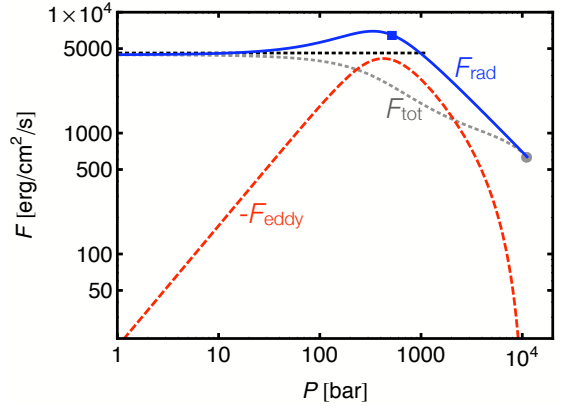}
\caption{Similar to \Fig{fig:Fluxdiff} except dissipation is included.  The total flux is no longer constant and the core flux is larger.  Both of these effects reduce the magnitude of $\Fe$ which can no longer offset  as much of $\Fr$.  Consequently $\Kzzmax$ is reduced to $900~{\rm cm^2/s}$.  The dissipation profile $\epsilon = \epsilon_o + \epsilon_{\rm buoy}$ includes a floor $\epsilon_o = 5 \times 10^{-5} ~{\rm erg/(g s)}$, that corresponds to $f_\epsilon = 0.01$, i.e. weaker dissipation near the RCB than in stratified regions.
 }
\label{fig:Fluxdiss}
\end{figure}

\subsection{Including Dissipation}\label{sec:diffdissres}
We now consider the effect of adding dissipation to our models with eddy diffusion.    The total flux $F$ will now increase with height due to dissipation.  The coupled \Eqs{eq:TPed}{eq:FPed} govern the steady state structure.  To understand the effect of dissipation on the turbulent heat flux, we return to the simpler case of spatially uniform $K_{zz}$. 

With dissipation we still find an upper limit to diffusion, $\Kzzmax$, for a given $T_1$ and $T_{\rm deep}$.  However $\Kzzmax$ declines with increasing dissipation.  \Fig{fig:Kzzdiss} shows this for  both a dissipation rate $\epsilon_o$ that is constant with height (\emph{solid blue curve}) and the full dissipation prescription  (\emph{dashed red curve}, see eq.\ [\ref{eq:diss}]). The full prescription includes $\epsilon_{\rm buoy}$, our estimate of the minimum dissipation due to stratified turbulence.  This additional dissipation further reduces  $\Kzzmax$.

We confirm that the dissipation-free estimates of $\Kzzmax$ in previous sections represent a conservative upper bound.  Lowering $\Kzzmax$ means that weaker turbulent diffusion will inflate the planet.  Admittedly, the cases shown in \Fig{fig:Kzzdiss} do not prove that all dissipation profiles will lower $\Kzzmax$.  However we investigated the effects of both spatially-varying $K_{zz}$ (as in \S\ref{sec:varyKzz}) and also different profiles of $\epsilon$.  In all cases adding dissipation reduced $\Kzzmax$ from the dissipation-free value.   

\subsubsection{How Dissipation Lowers $\Kzzmax$}
We explore the energetics of how dissipation lowers $\Kzzmax$.    
\Fig{fig:Fluxdiss} shows flux balance with dissipation, and can be compared to \Fig{fig:Fluxdiff}.  Notice that the peak value of $-\Fe$ now falls well short of $\Fr$ at the relevant pressure, $P_{\rm deep}$.
This is because the total flux 
\begin{equation} 
F = F_{\rm c} + \int^{P_{\rm c}}_P {\epsilon \over g} dP
\end{equation} 
now includes the integrated dissipation, causing $F$ to greatly exceed $F_{\rm c}$, the small loss of heat from the core.   From the RCB to $P_{\rm deep}$, $\Fr$ also increases with height.  This rise is not significantly affected by dissipation, with $\Fr \propto P_{\rm c}^{-6/7}$ along the pseudo-adiabat as  before.  The magnitude of $-\Fe = \Fr - F$ is smaller at $P_{\rm deep}$ because dissipation increases $F$ without comparably increasing $\Fr$.  

We find that $P_{\rm deep}$ (and also $\rho_{\rm deep}$) do not significantly change when we add dissipation.   Indeed the $T-P$ or $\nabla$ profiles for the solution in \Fig{fig:Fluxdiff} are indistinguishable from the stirred atmospheres in \Fig{fig:TPdiff}, except the RCB is not pushed as deep, ``only'' to  $P_{\rm c} \approx 11 ~{\rm kbar}$.  

The limiting value of $K_{zz} = 2|\Fe|/(\rho_{\rm deep} g)$ drops because the smaller eddy flux is not compensated by a lower $\rho_{\rm deep}$.  It seems possible that some dissipation profile could heat the atmosphere and lower $P_{\rm deep}$ and $\rho_{\rm deep}$ enough to increase $\Kzzmax$.  We did not find this to be the case.  One reason is that too much dissipation can affect the location of the RCB, a subject we address below.

For completeness we explain the flux balance at small pressure illustrated in \Fig{fig:Fluxdiss}.  The decline in $-\Fe \approx \rho g K_{zz}$ with height in isothermal regions again results from declining density.   However $\Fr$ does not decline towards low pressure (as was seen in \Fig{fig:Fluxdiff}),  because the total flux $F$ is larger with dissipation.    The fact that the escaping flux matches the flux from the radiative equilibrium solution (\emph{dotted black line}) is a coincidence (with some significance, see below).  This coincidence occurs because the dominant dissipation is $\epsilon \approx \epsilon_{\rm buoy}$.  Larger or smaller choices of dissipation would give a larger or smaller (respectively) net $F$ and escaping $\Fr$.  

This coincidence has some significance.   Downward eddy fluxes reduce the loss of heat from the core, as we have discussed extensively. However we now see that this loss is matched by the flux due to the dissipation of that turbulence --- provided our prescription for the minimum $\epsilon_{\rm buoy}$ is correct.  This replacement is intriguing, but does not alter our discussions of evolutionary consequences:  turbulent dissipation in radiative regions is powered not by interior heat, but by external means (such as forced atmospheric circulation). 

\subsubsection{The Effect of Dissipation on the RCB}
Ignoring dissipation, we obtained the limiting $\Kzzmax$ as $P_{\rm c} \rightarrow \infty$.  With dissipation, $\Kzzmax$ occurs at finite $P_{\rm c}$, provided there is dissipation at the RCB.  As noted above, the RCB occurs at $11$ kbar with $K_{zz} = \Kzzmax$ in \Fig{fig:Fluxdiss}.   Appendix \ref{sec:appRCBdiss} derives the relation between dissipation and the maximum $P_{\rm c}$ in \Eq{eq:disslim}.

We consider it physically desirable to restrict $P_{\rm c}$ to finite pressures.   Infinite $P_{\rm c}$ obviously violates some of our idealizations, notably the plane parallel and ideal gas approximations.  It is encouraging that dissipation  alters $P_{\rm c}$ without qualitatively changing the insights (namely $\Kzzmax$) gleaned from the non-dissipative model.

Restricting $P_{\rm c}$ to finite values was not crucial for the energetic balance arguments above.  Though $\Fc$ is larger for smaller $P_{\rm c}$, it is still too small to be the main factor that limits the eddy flux.  

To illustrate some of these points consider the case $\epsilon = \epsilon_{\rm buoy}$, i.e.\ with no floor, $\epsilon_o$, to the dissipation.  In this case there is no dissipation at the RCB, and we find that eddy diffusion can still push $P_{\rm c} \rightarrow \infty$.  Nevertheless $\epsilon_{\rm buoy}$ by itself does still lower $\Kzzmax$, by $\sim 1/3$.  This can be seen in the $\epsilon_o \rightarrow 0$ limits (i.e.\ the left of the plot) of \Fig{fig:Kzzdiss}.

\subsection{Varying the Opacity Law and EOS}\label{sec:opacity}
All of our plots and numerical estimates have used an opacity law $\kappa \propto P$ and an ideal gas EOS with $\delad = 2/7$.  We provide general scalings for many results to show the effect of varying these parameters.

Our qualitative results hold for all ``reasonable" choices of power law opacities and EOS.  As discussed in \S\ref{sec:rt}, reasonable means that $\nabla_\infty > \delad$ and $\alpha > -1$ so that radiative equilibrium solutions become convectively unstable at depth.  We tested our results with an alternate opacity law  $\kappa \propto T^2$  ($\alpha = 0$ and $\beta = 2$), that approximates H$^{-}$ opacities for $T > 2000$ K or dust grain opacities at colder temperatures.  Like $\kappa \propto P$, this law also has $\nabla_\infty = (1+\alpha)/(4-\beta) = 1/2 > \delad$. 

The behavior of $\Kzzmax$ --- shown in \Eq{eq:Kzzmax} --- is particularly important for interpreting our results.   The scaling with internal entropy is quite steep with $\Kzzmax \propto T_1^{10.5}$ and $\Kzzmax \propto T_1^7$ for our standard and alternate opacities, respectively.  Generally, the entropy  dependance becomes steeper for larger $\alpha$ (as in the above example) and also for a smaller $\delad$, but does not depend on $\beta$.   Recall that the burial of the turbulent heat flux into the convective interior can bring an atmosphere with $K_{zz} > \Kzzmax$ towards energetic equilibrium.  
A steeper dependance of $\Kzzmax$ on entropy means that less inflation is needed to enforce this equilbrium.  

The scaling of $\Kzzmax$ with $T_{\rm deep}$ controls how our inflation mechanism depends on the level of irradiation.  Also the development of thermal inversions will lower $T_{\rm deep}$ for a fixed level of irradiation.  We find $\Kzzmax \propto T_{\rm deep}^{-5.5}$ or $\Kzzmax \propto T_{\rm deep}^{-4}$, again for the standard and alternate opacities, respectively.  The $T_{\rm deep}$ dependance becomes more steeply negative for larger $\alpha$ (again the dominant effect in our example) and also for larger $\beta$ (less important in our example) and smaller $\delad$.

These scalings emphasize that a more detailed treatment of turbulent heat fluxes in hot Jupiters should include realistic opacities and equations of state.  Non-powerlaw behavior could have significant consequences.  For instance, \citet{Guillot_etal94} demonstrated the importance of an opacity window near $\sim 2000$ K.  This window can give rise to an isolated convective layer sandwiched between two radiative zones.  It would be interesting to consider how a downward turbulent heat flux would interact with such a region.  Future work that generalizes our self-similar approach can address these more detailed issues.  

\section{Comparison with Previous Work}\label{sec:comparison}
\subsection{Simulations of Atmospheric Circulation}\label{sec:dynamics}
Hydrodynamic simulations of hot Jupiter atmospheres have been studied in local \citep[][hereafter LG10]{Burkert2005,LiGoodman_10} and global \citep[e.g.][]{CooperShowman05,RauscherMenou10,Showman_etal09,Dobbs-DixonLin_08} models.  
A major goal of these studies is to determine the circulation induced by stellar irradiation.
We first discuss estimates of $K_{zz}$ from these simulations, and then address the question of whether radiatively forced turbulence extends throughout the radiative zone, as we have assumed.  Other sources of turbulence --- perhaps involving magnetic fields --- may also exist, but we do not analyze them here.

The global circulation models of \citet{Showman_etal09} estimate $K_{zz} \sim 10^{11} ~{\rm cm^2/s}$ at mbar pressures.  This does not contradict our constraint on upper atmospheric mixing from the efficiency of radiative forcing [\Eq{eq:Carnot}] when extrapolated to such low pressures.  Moreover at low pressures radiative losses can lower the eddy flux (see \S\ref{sec:Feddy}) and further weaken our  constraint on $K_{zz}$.
We caution that $K_{zz}$ estimates from GCMs are rough, since they only resolve relatively large scale flow patterns.  The \citet{Showman_etal09} $K_{zz}$ estimate arises from multiplying measured vertical speeds by the scale-height $H$.  While a useful guide to what is possible, this does not constitute a direct measurement of diffusion.

Local hydrodynamic simulations can better resolve turbulent flows, though as usual with Reynolds numbers far lower than reality.   The calculations of   LG10  find an effective turbulent viscosity of $\nu_t \sim 0.001$ --- $0.01 c_{\rm s} H \sim 10^{10}$ --- $10^{11}~{\rm cm^2/s}$.  It is unclear if this viscosity should be interpreted as a mixing coefficient.  Their simulation with $\nu_t = 0.015 c_{\rm s} H$ has an RMS vertical speed $w \sim 0.3 c_{\rm s}$.  Thus assuming $\nu_t \sim K_{zz}$ is not consistent with the simple estimate $K_{zz} \sim w \ell$ because it gives a length scale $\ell \sim H/20$, smaller than the grid spacing of $H/10$.  It should not be surprising that simple estimates based on 3D isotropic turbulence fail, given the organized structure in their 2D ``turbulent" state.

Moreover the forcing in LG10 was not by irradiation, but chosen to be large enough to drive  super-sonic flows despite artificial viscosities that are large for numerical reasons.  Thus attempting to interpret the Carnot efficiency of stellar irradiation may overextend their results.  Despite these caveats, if the LG10 simulations apply near the mbar weather layer, there is again no contradiction with \Eq{eq:Carnot}.

Global simulations indicate that the shear layer may not extend throughout the radiative zone.  For instance \citet{Showman_etal09} find that strong ($\sim $km/s) zonal winds terminate at $\sim 10$ bar.  
They argue that circulation stops because the planet is horizontally isothermal at these depths, removing the local forcing.  
However, hot Jupiter atmospheres have a large separation between the radiative timescales in the weather layer and the deep radiative layer. It is possible (and arguably likely) that unresolved or long timescale dynamics could push the shear layer even deeper.  This possibility should be explored in future modeling studies.

How far turbulence can extend below the shear layer is also uncertain.  LG10 force a shear layer of with $\sim 2 H$, but find that turbulence (or at least some disordered motion) extends throughout their box of size $\sim 5 H$ (see their Fig.\ 10).  Turbulence that extends  a full $5 H$ below a shear layer could thus extend to $P \sim e^5 \cdot 10~{\rm bar} \sim 1.5$ kbar.  These issues deserve further study, but it cannot be ruled out that the radiative zone is turbulent down to the RCB.

\subsection{Thermal Inversions via TiO Diffusion}\label{sec:S09}
We now provide a more detailed interpretation of our results in terms of the S09 constraints on the mixing required to loft TiO and create thermal inversions.  We emphasize that the constraints on $K_{zz}$, and especially on the depth dependance of $K_{zz}$ depends sensitively on whether or not there is a cold trap.  Wherever TiO condenses, one must consider the mixing of dust grains, not just molecules.

First consider the case where there is no cold trap, and TiO is always in the vapor phase.  In the S09 analysis, this is only possible for the most intensely irradiated planet, WASP 12b, in part because thermal inversions lower $T_{\rm deep}$ and favor condensation at depth.   The mixing of TiO vapor requires $K_{zz} \sim 10^7~{\rm cm^2/s}$ at $P \sim 1$ mbar, i.e.\ the height of the inversion where TiO is needed.  The specification of pressure level is important because the constraint on molecular diffusion scales as $K_{zz} \propto P^{-1}$ in an isothermal atmosphere.  Thus the constraint is weaker at depth.  In a model where the actual $K_{zz} \propto P^{-1}$ (as in \S\ref{sec:varyKzz}) we only require $K_{zz} \sim 10 ~{\rm cm^2/s}$ at the kbar RCB.  In this case our model predicts a minimal effect of the eddy flux on the planet's evolution, though the dissipation of turbulence above the RCB could still be significant.

We briefly summarize how this  $K_{zz} \sim 10^7~{\rm cm^2/s}$ limit and the depth dependance arises, and refer the reader to S09 for details.  With only molecular viscosity, i.e.\ collisions, the TiO scaleheight would be hydrostatic, and thus $\sim 30$ times smaller than the dominant $H_2$ species.  S09 conclude that the turbulent diffusion must exceed the molecular diffusion $D_{\rm TiO}$ by a factor $\sim 100$, due to the many ($\sim 14$)  scaleheights between $P \sim $ mbar and the kbar RCB.  Estimating $D_{\rm TiO}$ as usual --- the product of the mean free path and thermal speed --- gives an inverse scaling with density, and thus also with pressure in isothermal regions.   The numerical value of $D_{\rm TiO} \sim 10^5 ~{\rm cm^2/s}$ at $P \sim$ mbar, combined with the factor of $100$ excess needed for efficient turbulent mixing, gives the $K_{zz} \sim 10^7~{\rm cm^2/s}$ constraint.

Now consider the case where grains do condense somewhere in the radiative zone, which is true for most hot Jupiters (especially those with inversions).  The $K_{zz}$ required to loft TiO increases to $10^8$ --- $10^{11}~{\rm cm^2/s}$ depending on the size of grains and the depth of the cold trap.   A rough estimate of $K_{zz} \sim v_{\rm term} L$ follows from equating the diffusion timescale, $L^2/K_{zz}$, to the dust settling timescale $L/v_{\rm term}$, where $L$ is the depth of the cold trap and $v_{\rm term}$ is the grain's terminal speed.  While the terminal speed will increase with particle size, it does not depend on atmospheric density for the relevant viscous (i.e.\ Stokes') drag.  Thus unlike the case of molecular viscosity, the $K_{zz}$ required for mixing condensed grains does not decline with depth.

We can thus conclude that the $K_{zz}$ required to mix TiO and create thermal inversions in a hot Jupiter are likely excessive, provided TiO condenses somewhere in the radiative zone.   This follows from \Fig{fig:KzzS} which shows that $K_{zz} \sim 10^9 ~{\rm cm^2/s}$ appear off-scale.  The internal (specific) entropy of a planet would have to increase by $\gtrsim 2 k_{\rm B}/m\ps$ for a planet with $T_{\rm deep} \gtrsim 1500$ K, as is the case for all planets in S09.  It seems likely that such large entropies would overinflate a planet, based on studies such at AB06.  This is especially true because \Fig{fig:KzzS} neglects dissipation which can further inflate a planet, which can occur by lowering $\Kzzmax$ as shown in  \Fig{fig:Kzzdiss}.
 
However, we cannot firmly declare that the TiO hypothesis fails.  This is largely due to the approximate nature of our treatment of opacities and the equation of state.  A more conclusive analysis of the TiO hypothesis would require a detailed atmospheric model that includes the eddy fluxes and turbulent dissipation described in this work.  Such a study would also have to abandon the fixed flux bottom boundary condition used in most atmospheric models (including S09).  Instead the bottom boundary should be a fixed adiabat, chosen to match the observed radius.  This is subject as usual to assumptions about composition and presence of a core.  However allowing the flux to float is needed for a consistent determination of  the RCB.  This is crucial for understanding the coupled relationship between compositional mixing and cooling history.

\section{Conclusions}\label{sec:concl}
\subsection{Summary of Results}
We have investigated how forced turbulent mixing affects the energetic balance and structure of the stably stratified, radiative layers of hot Jupiters.    It is crucial to understand how the radiative layer matches onto the convective interior at the radiative-convective boundary (RCB).  This regulates the rate at which the planet cools and therefore controls the evolution of the planet's radius.  

Previous work has invoked turbulent eddy diffusion of molecular species (and dust). This mixing changes the opacity to provide better-fitting model spectra of transiting planets,  especially those that appear to have thermal inversions (S09). 
We find that turbulent mixing of this kind does not just redistribute chemical species but also significantly affects energetics.

Forced turbulent mixing in stable, radiative regions drives a downward flux of energy that pushes the RCB deeper in the atmosphere, lowering the planet's cooling rate.   We found an upper limit to the strength of turbulent diffusion, $\Kzzmax$, that can be achieved in steady-state.  Beyond this limit, the downward flux of energy will heat the convective interior and inflate the planet.  We did not directly model this entropy growth  because our model was steady state and did not include overshoot across the RCB.   The deep, marginally stable layer in our solutions would not strongly inhibit overshoot, so heat burial by this mechanism is a likely outcome.  Our solutions indicate that interior heating brings the planet back towards steady state, because higher entropy planets have larger $\Kzzmax$ (see \Fig{fig:KzzS}).  Our mechanism is thus a ``mechanical greenhouse effect" with the role of sunlight in the traditional greenhouse being played by forced turbulent mixing. 

For non-uniform turbulent mixing we find that our constraint on $\Kzzmax$ applies near the RCB.  More specifically this constraint applies at a pressure $P_{\rm deep}$, which lies below the deep isothermal region where the radiative layer is transitioning towards convective instability.  Our constraints on turbulence in the upper atmosphere are less stringent.  However, the downward flux of mechanical energy likely cannot exceed a small fraction (probably at the percent level) of the stellar irradiation if it is supplied by weather-layer winds.  If turbulence is too weak at the bottom of the radiative layer, it will not dredge up heavy molecular species --- particularly those that condense onto dust grains --- to serve as opacity sources near the observable photosphere.

Turbulence also deposits heat in the radiative atmosphere when it decays.  Non-turbulent sources of energy dissipation --- including non-linear wave breaking and ohmic dissipation --- also affect energetic balance.  We find that including energy dissipation reduces $\Kzzmax$, and thus makes it easier to inflate a planet for a given level of forced turbulence.

 We find a characteristic scale of $\Kzzmax \sim 10^3$ -- $10^5~{\rm cm^2/s}$ for typical hot Jupiter parameters, even ignoring dissipation.  This is orders of magnitude below values quoted in the literature of $10^7$ --- $10^{11}~{\rm cm^2/s}$ for the mixing of chemical species \citep{Spiegel_etal09,Zahnle_etal09}.  We caution that our quantitative results should be taken as illustrative, due to the approximations detailed in \S\ref{sec:model}.  
 
Thus when turbulence is important for the redistribution of opacity sources, it also has significant energetic consequences.   Our model represents a step towards more self-consistent modeling of exoplanet atmospheres.

\subsection{Applications and Extensions}
 
\citet{GuillotShowman_02} first noted that dissipation in radiative layers can slow planetary contraction by pushing the RCB to higher pressures.  They imagined dissipation concentrated in the upper regions of the atmosphere, sourced by insolation driven winds.  Other authors \citep[e.g.][]{Bodenheimer_etal03} have included the \citet{GuillotShowman_02} dissipation prescription to model exoplanet radii.

However we are unaware of previous works that include the mechanical flux of energy due to turbulence or waves --- our $\Fe$--- in radiative layers.  Incorporating this effect in detailed planetary evolution models  would require a more precise treatment than this exploratory study.  Notably it would require a global model with a realistic equation of state and opacity, not the power laws considered here.

In this paper we treat diffusion and dissipation by forced turbulence as free parameters to allow a general analysis.   This approach is justified since we currently lack a detailed understanding of these processes.  Thus our model can be used to test how any specific model for turbulence and/or energy dissipation affects the structure and evolution of the planet.   For instance, the work of \cite{BatyginStevenson_10} consider ohmic dissipation in radiative and convective regions, but do not recompute the effect of the dissipation on the structure of the radiative layer.  Our point is not to critique, but to emphasize that  a more consistent treatment of energetics may make it easier to inflate a planet --- by any number of mechanisms.

We also hope to include the effects of our study in detailed 3D radiation-hydrodynamical simulations of exoplanet atmospheres.  
This would involve adding sub-grid physical prescriptions for eddy fluxes  to global circulation models (GCM) such as those by \cite{Showman_etal09, RauscherMenou10}  and/or non-hydrostatic simulations \citep{Dobbs-Dixon_10}.  \citet{Goodman_09} discusses the need to include explicit sources of dissipation. \citet{LiGoodman_10} present a first step towards sub-grid modeling of  turbulence due to Kelvin-Helmholtz instabilities.  Breaking of vertically propagating gravity waves represent another possible source of turbulence \citep{ShowmanChoMenou09}.  When all these effects are taken into account, the inflated radii of transiting planets may not be surprising after all.

We have focused on hot Jupiters, but the physics we describe in principle applies to other atmospheres.  Our limits  on eddy diffusion become less severe as objects are more weakly irradiated (see \Fig{fig:KzzS}).  Thus our mechanism would not inflate longer period gas giants, including Jupiter and Saturn.  More intrinsically luminous objects like brown dwarfs will similarly be less affected.  The expanding inventory of transiting exoplanets, at wider separations, is an excellent test of radius evolution models.

Finally we comment on a possible connection to the atmosphere of Venus.  Venus has a marginally stable pseudo-adiabat beneath a thick cloud deck \citep{Schubert_etal80}.  The clouds shield sunlight from warming the surface of Venus.  Because of this, Venus' atmosphere would be nearly isothermal were it not for some poorly understood mechanical stirring process.  Our well-stirred hot Jupiter atmospheres also have marginally stable pseudo-adiabats at depth.  Perhaps Venus is a very-well-stirred analog of a hot Jupiter.  Obvious differences exist, for instance the non-negligible fraction of sunlight that reaches the Venusian surface versus the radiant flux escaping the core of hot Jupiters.  Pursuing analogies such as these should improve our understanding of worlds near and far. 

\acknowledgments
ANY and JLM thank David Spiegel for extensive discussions and Peter Goldreich for wise counsel.  ANY thanks William Hubbard and  Roger Yelle for insights into planetary cooling  and eddy diffusion, respectively.  JLM thanks Martin Pessah and Shane Davis for helpful discussions.  JLM and ANY thank the Institute for Advanced Study for hosting them during the early stages of this work.  This research was supported in part by the National Science Foundation under Grant No. PHY05-51164 for ANY to visit the KITP.  ANY thanks  Travis Barman, Lars Bildsten, Brad Hansen and Emily Rauscher for useful feedback during the KITP program ``The Theory and Observation of Exoplanets."  We thank the anonymous referee, Kristen Menou, David Spiegel, Tristan Guillot and David Stevenson for comments that improved the submitted manuscript.


\appendix
\section{Zero Flux Solutions}\label{sec:zeroflux}
In addition to the solution technique described in \S\ref{sec:tech}, we also used a more specialized technique  that applies to a reduced model with no dissipation and a net flux $F=0$.  As we show in  \S\ref{sec:constKzz},  this is an interesting case because the downward flux of heat can drive the core flux, $\Fc \rightarrow 0$, and the flux remains zero in the absence of dissipation.  This alternate technique allows us to check our results.  It allows integration from the top down, compared to the bottom up integration from the RCB (which recedes to infinite pressure for the zero flux solution).   We do not need to specify the parameter $\psi_{\rm c}$ (eq.\ [\ref{eq:psi}]), which diverges for these zero flux solutions.  With the reduced model we can analytically explore solution properties.  We will do this below to explain why we cannot obtain solutions with $\zeta \la -1.5$ in \S\ref{sec:varyKzz}.

Ignoring dissipation and setting $F=0$ we can rearrange \Eq{eq:TPed} as
\begin{equation} \label{eq:Fzero}
{d T \over d P} = \delad{T \over P} \left(1+ {\delad k_{\rm rad} T \over \rho g K_{zz} P }\right)^{-1} \, .
\end{equation} 
We can integrate from $P=0$ with the boundary condition $T(0) = T_{\rm deep}$.  The solutions approach an adiabatic profile $T \propto P^{\delad}$ at large pressure (shown below).  One could find solutions by integrating with various $K_{zz}$ values, iterating until you land on a desired adiabat.  Instead we again take a self-similar approach, non-dimensionalizing the pressure using $K_{zz}$ and the opacity law.  We choose a physical scale for the pressure by matching the self-similar solution onto a chosen adiabat.   This fixes the value of $K_{zz}$ (at a reference pressure if $K_{zz}$ is not uniform).  The special value of $K_{zz}$ that gives a zero flux solution for a given adiabat is the limiting $\Kzzmax$ discussed in \S\ref{sec:constKzz}.

We now show that the solution to \Eq{eq:Fzero} becomes adiabatic at large pressure, at least if $K_{zz}$ is constant.   If the final term in parenthesis vanishes we have $\nabla \rightarrow \delad$.  We can show that an adiabatic solution is consistent by assuming $T \propto P^{\delad}$ and then confirming that the final term
\begin{equation} \label{eq:scaletest}
 {\delad k_{\rm rad} T \over \rho g K_{zz} P } \propto {T^{5-\beta} \over P^{2+\alpha}} \propto P^{(2+\alpha)(\delad/\nabla_\infty' -1)} \ll 1
\end{equation} 
as $P\rightarrow \infty$.  With $\nabla_\infty' \equiv (2+\alpha)/(5 -\beta)$ we can show that the exponent in the last proportionality of \Eq{eq:scaletest} is indeed negative.  This is because $\nabla_\infty' > \nabla_\infty$ (which is generally true, for our opacity $\nabla_\infty' = 3/5 > 1/2$) and because  reasonable opacities have $\alpha > -1$ and $\nabla_\infty > \delad$ (see \S\ref{sec:model}).   Thus a deep adiabat is a consistent solution at large pressure, for constant $K_{zz}$.

Now consider a spatially varying $K_{zz} \propto P^\zeta$.  In this case the assumption of an adiabat at depth gives
\begin{equation} 
{\nabla \over \delad} \rightarrow {1 \over 1 + c_\zeta P^{e_\zeta}}
\end{equation} 
as $P \rightarrow \infty$, where $c_\zeta$ is constant and $e_\zeta = \delad(5-\beta) - (2+\alpha+ \zeta)$.  Consistency (i.e.\   $\nabla \rightarrow \delad$) requires that $e_\zeta < 0$ or
$\zeta > -(2+\alpha) + \delad(5-\beta) = -11/7 \approx -1.57$.  More negative values of $\zeta$ do not approach an adiabat even at infinite pressure.

This explains the behavior of $\zeta \la -1.4$ solutions in \S\ref{sec:varyKzz}.  However we should not over-interpret the physical significance of this result since it may merely reflect a limitation of the self-similar approach.  More study is needed of what happens when a large mechanical flux of energy enters the top of the atmosphere, but turbulent heat burial is very weak near the RCB (which is the case for negative $\zeta$).  For the time being, we thus consider the irradiation efficiency constraint in \Eq{eq:Carnot} as our best limit on turbulent mixing in the upper atmosphere.

\section{Maximum Dissipation at RCB}\label{sec:appRCBdiss}
This appendix derives how dissipation at the RCB restricts the RCB depth $P_{\rm c}$.  We use the consistency requirement that $d \nabla/dP \geq 0$ at the RCB.  This states that the solution will indeed be convectively stable above the RCB.  We rearrange \Eq{eq:TPed} to give
\begin{equation} 
{\nabla \over \delad} = {F + F_{\rm iso} \over \delad k_{\rm rad} T/P + F_{\rm iso}}
\end{equation}
and at the RCB (denoted by the `c' subscript) we have $F = F_{\rm c} = \delad (k_{\rm rad} T/P)_{\rm c}$.  The gradient of $\nabla$ at the RCB is
\begin{equation} 
{d \over dP}\left({\nabla \over \delad}\right)_{\rm c} = {1 \over F_{\rm c} + F_{\rm iso,c}} {d \over dP}\left[F - {\delad k_{\rm rad} T \over P} \right]_{\rm c} 
\end{equation} 
where the derivatives of $F_{\rm iso}$ cancel.  Thus the limit on dissipation at the RCB, using \Eq{eq:FP} and requiring $(d \nabla/dP)_{\rm c} \geq 0$, is
\begin{equation} \label{eq:disslim}
\epsilon_{\rm c} \leq (1+\alpha) \left(1-{\delad \over \nabla_\infty}\right){F_{\rm c} g \over P_{\rm c}} =  {6 F_{\rm c} g \over 7 P_{\rm c}}\, .
\end{equation} 
Since $F_{\rm c}/P_{\rm c} \propto P_{\rm c}^{-11/7}$,  the maximum depth of the RCB declines with increasing dissipation at the RCB.

This result is very useful in finding $\Kzzmax$ values with dissipation.     Without dissipation we could easily find $\Kzzmax$ by increasing $\psi_{\rm c}$ (eq.\ [\ref{eq:psi}]) to arbitrarily large values (which results in \Fig{fig:Kzzdiff}), or by using  the zero flux solutions of appendix \ref{sec:zeroflux}.  Instead we add a twist to the self similar technique of \S\ref{sec:tech}.   We set the dissipation at the RCB to the limiting value of \Eq{eq:disslim}.  We further assume the dissipation is of the form of \Eq{eq:floor}  so that the strength of the dissipation is set by the dimensionless $f_\epsilon$.  We have thus coupled dissipation to $K_{zz}$ mathematically.  (It doesn't matter if they are unrelated physically, since $f_\epsilon$ can be adjusted to give any desired level of dissipation.)  Then we can solve equations (\ref{eq:psi}), (\ref{eq:Ndeep}), (\ref{eq:floor}) and  (\ref{eq:disslim}) for 
\begin{equation} \label{eq:psidiss}
\psi_{\rm c} = {6 T_{\rm deep} \over 7 f_\epsilon \delad T_{\rm c}}\, .
\end{equation} 
Since $T_{\rm c}$ is not known until we obtain a solution, we use an iteration procedure: guess a value for $T_{\rm c}$ (but start with something too low), use the estimate of $\psi_{\rm c}$ from \Eq{eq:psidiss} to integrate the model \Eqs{eq:TPed}{eq:FPed}, use the resulting $T_{\rm c}$ to refine $\psi_{\rm c}$ and repeat.  Though a bit convoluted, this procedure converges. 

Finally note that our constraint on dissipation at the RCB does not ensure stability at all $P < P_{\rm c}$.  Assume that the inequality in \Eq{eq:disslim} holds so that there is a radiative layer above the RCB.  Dissipation could still create another convective layer at greater height.  However our self-similar solutions cannot include such structures.  Thus we leave it to a future work to consider these more complicated scenarios and their effects on planetary evolution.

\end{document}